\documentclass[
          twoside=false,
          twocolumn=true,
          a4paper,
          10pt]{scrartcl}

\usepackage[top=1.5cm, bottom=2.5cm, left=1.5cm, right=1.5cm]{geometry}

\usepackage{ifpdf}

\ifpdf
   \usepackage[pdftex]{graphicx}
   \DeclareGraphicsExtensions{.pdf,.jpeg,.png}
\else
  \usepackage[dvips]{graphicx}
  \DeclareGraphicsExtensions{.eps}
\fi

\usepackage[caption=false,font=normalsize,labelfont=sf,textfont=sf]{subfig}

\usepackage{flushend}

\usepackage{multirow}

\begin{document}

\title{Evaluating Connection Resilience \\ for the Overlay Network Kademlia}

\author{
  Henner Heck, Olga Kieselmann and Arno Wacker \\
 \texttt{\{henner.heck$|$olga.kieselmann$|$arno.wacker\}@uni-kassel.de}
}

\date{}

\maketitle

\begin{abstract}

Kademlia is a decentralized overlay network, up to now mainly used for highly scalable file sharing applications.
Due to its distributed nature, it is free from single points of failure.
Communication can happen over redundant network paths, which makes information distribution with Kademlia resilient against failing nodes and attacks.
This makes it applicable to more scenarios than Internet file sharing.
%One of them are distributed Cyber-physical Systems, a concept becoming increasingly important in the context of Industry~4.0 automation.
In this paper, we simulate Kademlia networks with varying parameters and analyze the number of node-disjoint paths in the network, and thereby the network connectivity.
A high network connectivity is required for communication and system-wide adaptation even when some nodes or communication channels fail or get compromised by an attacker.
With our results, we show the influence of these parameters on the connectivity and, therefore, the resilience against failing nodes and communication channels.

\end{abstract}

\section{Introduction}
\label{conn:sec:introduction}

Kademlia~\cite{maymounkov2002kademlia} is a well known distributed overlay network which is mainly used for Internet file sharing, e.g., BitTorrent~\cite{loewenstern2008bep}.
It has a decentralized structure with redundant communication paths eliminating single points of failure.
This makes it suitable for other research fields, e.g., the emerging Industry~4.0 context.
In Industry~4.0, distributed Cyber-Physical Systems (CPS), consisting of multiple networked nodes, are expected to improve automated industrial processes significantly \cite{lee2015cyber}.
The networked nodes of a CPS interact with their physical environment using sensors and actuators, and store information about its state and development.
Two examples for distributed CPS are a smart camera network (SCN) and a network intrusion detection system (IDS).
In an SCN, multiple networked cameras collaborate in surveiling and tracking developments in an observed area.
An IDS secures cooperate networks with several branches by collaboratively detecting distributed attacks.

A common requirement for all these systems is the ability to exchange information between nodes.
%Since nodes of the system can fail, e.g.~from technical problems or attacks, the communication structure must be resilient enough to tolerate failures to a certain degree.
%In both given examples, the system must be able to adapt to complex situations in the physical environment going beyond the scope of a single node.
%A necessary feature for this is the information exchange about the physical environment between the nodes.
This information must be exchanged via communication channels, which can be either direct or indirect via other nodes.
However, we must consider that nodes or communication channels might fail.
%Additionally, there exist a number of security weaknesses as described in, e.g.,~\cite{broy2012cyber,Cardenas2008,eckert2013, krueger2013, sander2013}.
Since some nodes might be publicly accessible, we must consider that they can fail due to an attack.
To still achieve reliable communication, we require redundant communication channels for resilient inter-node communication \cite{heck2016multi}.
More precisely, to tolerate failing nodes, there must be multiple node-disjoint communication paths through the network for any node pair.
The minimum number of node-disjoint paths for any node pair in a network is the \emph{network connectivity}.

The main contribution of this paper is a thorough evaluation of the connectivity of the Kademlia overlay network, yielding the resilience of the network against node failures and disturbed communication channels.

The rest of this paper is organized as follows: First, we discuss related research about overlay network connectivity in Section~\ref{conn:sec:relatedwork} and present our assumptions in Section~\ref{conn:sec:systemmodel}. After that, we briefly describe in Section~\ref{conn:sec:connectivity} the Kademlia protocol and the mathematical foundations for computing the network connectivity. Based on this, we present  and discuss the results of our connectivity measurements in Section~\ref{conn:sec:evaluation}. Finally, we conclude our paper in Section~\ref{conn:sec:conclusion} with a brief summary and provide an outlook on future research.

% RW
% SM
% CONNECTIVITY
% EVA
% DISCUSSION
% CONCLUSION

\section{Related Work}
\label{conn:sec:relatedwork}

Kademlia and overlay networks in general have been studied extensively in the scientific literature. A survey about research on robust peer-to-peer networks from 2006 \cite{risson2006survey} already lists several hundred references. Another survey from 2011 with focus on security aspects, reaches close to a hundred references \cite{urdaneta2011survey}.
Despite the large amount of publications in general, the global network connectivity of Kademlia has not been thoroughly evaluated. We limit our discussion of related work to literature with relevance for connectivity of structured overlay networks built with Kademlia or it's descendants.

In \cite{kovacevic2008towards}, the authors simulate Kademlia networks and apply churn (joining/leaving of nodes) to evaluate resilience. While the basic premise is similar to ours, they measure response times and number of message hops, not network connectivity.
In \cite{jimenez2009connectivity}, the authors insert nodes into a real-world BitTorrent network.
The main focus of this paper is on connectivity problems within the network caused by technical obstacles such as firewalls and network address translation (NAT).
The authors analyze connectivity properties of small groups of nodes. They do not measure the network-wide connectivity.
Similarly, the authors of \cite{crosby2007analysis} insert nodes into real-world overlay networks built by the BitTorrent protocol to measure round trip times and message rates for resource lookups. Additionally, they measure ``connectivity artifacts'' and ``communication locality''. Artifacts emerge from nodes making contact with the author's nodes, but cannot be contacted by them. As in \cite{jimenez2009connectivity}, the authors conclude that such a behaviour is most likely caused by firewalls and NAT.
The communication locality measurements show to what degree nodes preferably communicate with other nodes that, according to the protocol's definition of node distance, are near to them.
While both properties are related to the network connectivity, it is not measured or derived.
The authors of \cite{salah2013capturing} present a crawling software for capturing connectivity graphs of networks built by the KAD protocol, a descendant of Kademlia.
They insert specially modified crawling nodes into real-world networks to contact other nodes and dump the contents of their routing tables.
Those tables are then used to create connectivity graphs of the networks. In \cite{salah2014characterizing}, the same authors characterize the resilience of those connectivity graphs and of other graphs resulting from simulations.
While their goal is similar to ours, their approach is of statistical nature and does not calculate the network connectivity.
In \cite{baumgart2007s}, the authors propose different measures to make Kademlia networks more resilient towards malicious nodes. One of those measures is the use of node-disjoint paths for lookup procedures. The authors measure success rates for lookup procedures using different numbers of disjoint paths. Their simulations imply that a certain average level of connectivity is present in a network, but they do not measure the actual connectivity.

In contrast, our main goal is to determine the network connectivity of Kademlia in dependence of its parameters. Some of the related work, e.g., \cite{baumgart2007s}, even rely on a given network connectivity, but it was determined neither analytically nor experimentally before.

\section{System Model}
\label{conn:sec:systemmodel}

%The concept for our system model and attacker model, as well as the evaluation scenarios, were initially presented in \cite{hahner2013concept}. We further published a comprehensive description from a security point of view in \cite{heck2016multi}. Since the topic of this paper is on network robustness for distributed CPS, we made minor changes to emphasize this aspect.
We consider a distributed system consisting of multiple networked \emph{nodes}.
The basic functioning of one node is not dependent on the functioning of others.
%The self-organization and specifically the self-adaptation at runtime is based on the Observer/Controller architecture (e.g., \cite{tomforde2011observation, hahner2013concept}).
The nodes exchange information for collaboration purposes, depending on the systems specific purpose and implementation.
%To achieve sufficient observation and adaptation and to choose appropriate means of control, multiple nodes must cooperate.
%This requires the exchange of information among the nodes.
%Therefore, while each node has its own local task, it also communicates and collaborates with other nodes towards a global system goal.
%The communication happens loosely coupled via messaging.
Each node is able to communicate with any other node, either directly or indirectly via others.
%The information exchange can happen in a ``push'' manner, where a node disseminates information it gains from observing the physical environment into the network, or in a ``pull'' manner, where a node requests information from other nodes.
The communication structure is organized by the Kademlia overlay network (cf.\ Section \ref{conn:subsec:kademlia}).
%The communication structure of the CPS is organized as a structured overlay network, i.e., based on Kademlia (cf.\ Section \ref{conn:subsec:kademlia}).
%In case of changes in the physical environment that go beyond the scope of a single node, multiple nodes or even all of them might need to collaborate to achieve sufficient observation and adaptation, and to choose appropriate means of control.

We assume the presence of an \emph{attacker} with the goal of disturbing, disabling or controlling nodes and communication channels.
We call a node which has been successfully attacked a \emph{compromised node}.
There are several other causes exhibiting the same effect as a compromised node, e.g., maintenance, failures from defects, or other disturbances like power outages.
Without additional measures, these are indistinguishable from an attack.
If an attacker has compromised a node, we assume that she is able to fully impersonate the node towards the rest of the system.
Therefore, an attacker can disseminate information into the network as a legitimate part of the system and also deny requests coming from other nodes and, thus, hinder or prevent information exchange.
Communication between two nodes is not always direct, so other nodes can be necessary for message transfer.
%Therefore, a compromised node includes the case of compromised communication channels.
Additionally, we assume that the attacker can subvert at most $a$ arbitrary nodes at any time.
With regard to communication channels, we assume an attacker or other causes can disturb the channel causing message loss.
This leads to a certain percentage of sent messages not reaching their destination.

%As the integration of computational and physical processes is a defining property of a cyber-physical system, attacks on the \emph{physical environment} can affect the nodes and transitively their communication as well. Such an attack would generally aim to deceive a node in its observation of the physical environment by modifying sensor input.\\
%Our evaluation focus is on \emph{compromised nodes} and their effect on the \emph{network connectivity} of the CPS.

\section{Connectivity}
\label{conn:sec:connectivity}

In this section, first, we present the properties and mechanisms of Kademlia important for routing and contact management.
To analyze the network connectivity, we introduce the mathematical foundations to transfer the network structure of Kademlia into the domain of graph theory by creating a connectivity graph.
Next, we describe the mathematical algorithms and necessary graph transformations for calculating the graph connectivity. Finally, we use the mathematical foundations to define the resilience of the communication network.
%We then transform that graph to be able to measure its connectivity using a maximum flow approach.

\subsection{Kademlia}
\label{conn:subsec:kademlia}

%Enabling information exchange between nodes in a very large distributed system is a challenging task. Centralist solutions, be it for transport or storage of information, often present a bottleneck for the system performance since they do not scale well. Beyond that, a centralist solution is also a security and reliability risk, as it presents a single point of failure. In case such a solution fails, the whole system fails.

With Kademlia, each node and each stored data object is identified by a numerical \emph{id} with the fixed bit-length $b$. These identifiers are generated from a node's network address or the data object respectively, using a cryptographically secure hash function with the goal of equal distribution of identifiers in the identifier space.
Each node maintains a routing table with identifiers and network addresses of other nodes, its so-called \emph{contacts}.
The routing table consists of $b$ so-called $k$-buckets to store the contacts of the node.
The buckets are indexed from $0$ to $b-1$, and the contacts are distributed into these buckets depending on the distance of their identifiers $\mathit{id}_i$ and the node's id.
For this, the distance between two identifiers is computed using the XOR metric, meaning that for two identifiers $\mathit{id}_a$ and $\mathit{id}_b$ the distance is $\mathit{dist}(\mathit{id}_a,\mathit{id}_b)=\mathit{id}_a \oplus \mathit{id}_b$, interpreted as an integer value.
The buckets are populated with those contacts $\mathit{id}_i$ fulfilling the condition $2^i \leq  \mathit{dist}(\mathit{id},\mathit{id}_i) < 2^{i+1}$, with $i$ being the bucket index.
This means that the bucket with the highest index covers half of the id space, the next lower bucket a quarter of the id space, and so on.
The maximum number of contacts stored in one bucket is $k$.
%To avoid confusion with the resilience value $k$ we rename the maximum bucket size $k$ to $k_{K}$ for the remainder of this paper.
Next to $b$ and $k$, another defining property of a Kademlia setup is the request parallelism $\alpha$, which determines how many contacts are queried in parallel when a node wants to either locate another node or retrieve/store a data object. Greater values can speed up the operation, while at the same time increasing the resulting network load.
The staleness limit $s$ determines how often in a row the communication with a contact must fail, so that it is considered stale and removed from the routing table. Greater values of $s$ delay the removal of actually stale nodes by waiting for more failed communication attempts, while small values might lead to a frequent removal of non-stale nodes due to a disturbed communication channel.
The Kademlia authors set the default values $b=160$, $k=20$, $\alpha=3$, and $s=5$.

The nodes of a Kademlia network can locate resources (other nodes, data objects) by means of their identifiers. Given a target identifier, a node queries $\alpha$ nodes from its routing table closest to that identifier. Those, in turn, answer with their own list of closest nodes, which can then be used in new queries. This way, the requesting node iteratively gets closer to the target identifier.
This process ends when a number of $k$ nodes have been successfully contacted, or no more progress is made in getting closer to the target identifier. The purpose of a lookup procedure is to locate a node or data object, the purpose of a dissemination procedure is to locate appropriate nodes for storing a data object.

% Now some properties
%In addition to acting as a building block for a larger self-organizing system, we find it remarkable that Kademlia itself also exhibits \emph{self-*} properties. Depending on its own identifier and that of other nodes, each node will build a different routing table, i.e., perform \emph{self-configuration}.
%Emerging properties for the overlay network are its connectivity and a small number of hops necessary for locating nodes or data objects.
%The protocol enables nodes to detect stale contacts in their routing table and replace them to restore connectivity, thereby exhibiting \emph{self-healing} behaviour.
%On replacing a stale node from the routing table, the most recently seen node from a list of possible replacements is chosen. This is done to optimize the chance of a non-stale replacement node. By doing so, the network continuously \emph{self-optimizes} its routing.

\subsection{Connectivity Graph}
\label{conn:subsec:connectivity_graph}

The representation of the network structure as a connectivity graph enables the application of concepts and algorithms from graph theory to analyze properties of the network.
The connectivity graph $D(V,E)$, with the vertices $V$ and edges $E$, is a directed graph representation of the nodes and their routing tables.
Each vertex from the connectivity graph represents a distinct node from the network. Hence, the number of vertices equals the number of network nodes.
To construct the connectivity graph, we add edges the graph according to the routing table of Kademlia.
For each node pair $\mathit{id}_i$, $\mathit{id}_j$ represented in the graph by vertices $v$ and $w$ respectively, we insert the directed edge $(v,w)$ into the set of edges $E$ if and only if node $\mathit{id}_j$ is present in the routing table of $\mathit{id}_i$.

Generally, in network graphs, a capacity value is often assigned to the edges for expressing the communication bandwidth between nodes. This is not a necessity for connectivity graphs, since the existence of the edges is enough to indicate a connection between nodes. However, since it is necessary for later steps, we assign a capacity of $1$ to each edge.

\subsection{Vertex Connectivity for Vertex Pairs}
\label{conn:subsec:vertex_connectivity_pair}

A directed edge in the connectivity graph $D(V,E)$ can be interpreted as an one-way water pipe.
The maximum amount of water able to flow through the pipe per time unit is modeled by the edge capacity.
%The maximum flow between two vertices
The \emph{maximum flow} between two vertices $v$ and $w$ is the sum of the capacities of the \emph{minimum edge cut}.
This is the set of edges with the smallest total capacity whose removal would cut off any flow from $v$ to $w$.
In other words, the minimum edge cut is the bottle neck which determines the maximum possible flow $v$ to $w$.

% With all capacities in our connectivity graph being $1$, each edge in the minimum edge cut contributes $1$ to the maximum flow, which makes the maximum flow equal to the number of edges in the minimum edge cut.

%As shown in \emph{Menger's theorem} \cite{menger1927allgemeinen}, which is applicable for directed graphs \cite{beineke2013topics}, the number of edges in the minimum edge cut for $v$ and $w$ is equal to the maximum number of pairwise edge-independent paths from $v$ to $w$. This allows us to compute the number of pairwise edge-disjoint paths from one node to another in our connectivity graph by computing the maximum flow.

Analog to the minimum edge cut for two vertices $v$ and $w$, the \emph{minimum vertex cut} is the minimum number of vertices whose removal cuts all paths from $v$ to $w$.
The order of the minimum vertex cut is called the \emph{vertex connectivity} from $v$ to $w$, i.e., $\kappa(v,w)$.
Menger's theorem for directed graphs states that for the two non-adjacent vertices $v$ and $w$ the vertex connectivity is equal to the maximum number of pairwise vertex-disjoint paths from $v$ to $w$ \cite{menger1927allgemeinen,beineke2013topics_ch0}.
This number correlates directly with the communication resilience (cf.~Section \ref{conn:subsec:resilience}).
Therefore, to evaluate the resilience, we need to calculate the vertex connectivity.

%For adjacent vertices $v$ and $w$ we use the definition for $\kappa(v,w)$ from \cite{beineke2013topics_ch12}.
%Here, the vertex connectivity of two vertices $v$ and $w$ is defined as the graph connectivity (cf.\ next section) of the graph without the edge $(v,w)$.
%$\kappa(v,w) = \kappa(D(V,E-(v,w))) + 1$.

There are multiple algorithms to compute the maximum flow/minimum edge cut between any two vertices in a graph.
However, in general, the vertex connectivity does not correspond to the maximum flow/minimum edge cut.
To bridge the gap from computing the maximum flow/minimum edge cut to computing the vertex connectivity, we apply Even's algorithm (e.g., \cite{even1973algorithmic, even1975algorithm, even1979graph, even2011graph}).
It transforms the connectivity graph $D(V,E)$ such that the maximum flow between two non-adjacent vertices is equal to their vertex connectivity.
This allows the application of maximum flow algorithms to calculate the vertex connectivity.
Even's graph transformation is applied on the original connectivity graph $D(V,E)$ consisting of $n$ vertices and $m$ edges. We assume that $D(V,E)$ has neither self-loops nor parallel edges.
The problem transformation is done by applying the following steps to each vertex of $D(V,E)$:

\begin{itemize}
\item Let $v$ be a vertex of the directed graph $D(V,E)$ with the incoming degree of $d_{\mathit{in},v}$ and outgoing degree of $d_{\mathit{out},v}$.
\item Split $v$ into the two vertices $v'$ (incoming vertex) and $v''$ (outgoing vertex).
\item All incoming edges of $v$ point to $v'$, so that it has the incoming degree $d_{\mathit{in},v}$.
\item Make all outgoing edges of $v$ originate from $v''$, so that it has the outgoing degree $d_{\mathit{out},v}$.
\item Insert the edge $(v',v'')$ with capacity $1$. Now the outgoing degree of $v'$ and the incoming degree of $v''$ are $1$.
\end{itemize}

\begin{figure}
\subfloat[Original Network Graph $D$.]{\includegraphics[width=\columnwidth]{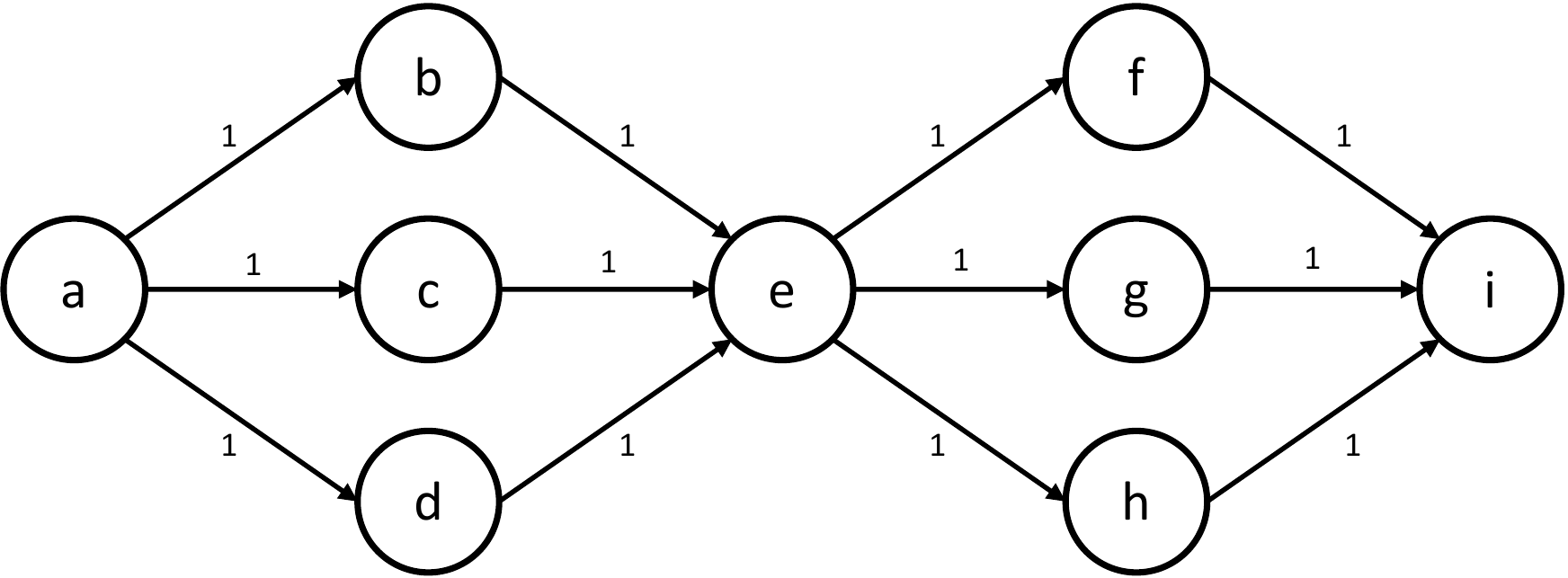}} \qquad \qquad \qquad
\subfloat[Transformed Network Graph $D'$.]{\includegraphics[width=\columnwidth]{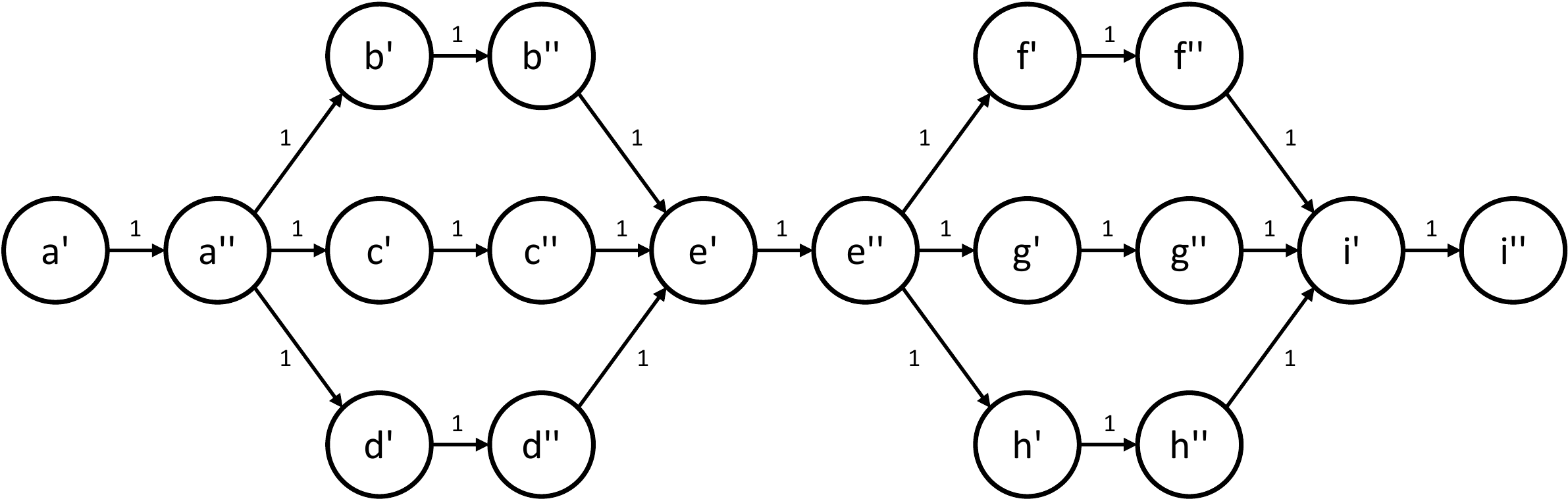}}
\caption{Example transformation for Even's algorithm. From vertex $a$ to vertex $i$, the connectivity graph in (a) shows a maximum flow of $3$ and a vertex connectivity $\kappa(a,i) = 1$. For $a''$ and $i'$ in the transformed graph $D'$ in (b), the maximum flow equals the vertex connectivity of $1$.}
\label{even_graphs}
\end{figure}

The resulting graph $D'(V',E')$ has $2n$ vertices and $m+n$ edges and can be used to calculate the vertex connectivity by applying a max flow algorithm.
An example for such a graph transformation is shown in Figure \ref{even_graphs}.

\subsection{Vertex Connectivity for Graphs}
\label{conn:subsec:vertex_connectivity_graph}

The vertex connectivity of a graph $D(V,E)$ is the minimum of the vertex connectivities of all pairs of distinct non-adjacent vertices in the graph, i.e.,
\begin{equation}
\kappa(D)=\mathit{min}(\kappa(v,w)) : v \neq w \wedge (v,w) \notin E \wedge v,w \in V.
\end{equation}
If $D(V,E)$ is not a complete graph, we determine the vertex connectivity $\kappa(v,w)$ for a pair of non-adjacent vertices $v$ and $w$ by computing the maximum flow from outgoing vertex $v''$ to the incoming vertex $w'$ in the transformed graph $D'(V',E')$. Therefore, the vertex connectivity $\kappa(D)$ for the whole graph can be determined by finding the minimum of the maximum flows between all pairs of outgoing and incoming vertices in the transformed graph $D'(V',E')$.
If $D(V,E)$ is complete, meaning that any vertex is adjacent to any other vertex, the vertex connectivity is the number of vertices in the graph $n$ minus one \cite{beineke2013topics_ch0}.

To find the minimum of the maximum flows for a transformed directed graph $D'(V',E')$ with $2n$ vertices, it is generally necessary to compute the maximum flow for all $n(n-1)$ distinct pairs of outgoing/incoming vertices. This makes the time complexity in terms of maximum flow computations $\mathcal{O}(n^2)$.
To find the minimum of the maximum flows for a transformed undirected graph $G'(V',E')$ with $2n$ vertices, it is sufficient to compute the maximum flow for $n-1$ distinct pairs of outgoing/incoming vertices \cite{gomory1961multi}. This makes the time complexity in terms of maximum flow computations $\mathcal{O}(n)$.

% nach der Mathematik
\subsection{Resilience}
\label{conn:subsec:resilience}

%Hence, to tolerate $s-1$ compromised nodes, there must be at least $s$ node-disjoint communication paths through the network for any node pair. In mathematical terms, this means that the network connectivity must be at least $s$ -- we, therefore, call $s$ the security level of the CPS communication network.
% To fulfill this requirement the CPS network needs to possess a sufficient resilience towards attacks.
We call a network that can function properly even when a number of $r$ nodes have been compromised an $r$-resilient network.
This means that with up to $r$ compromised nodes a path still exists between any pair of nodes.
As stated in our system model (cf.\ Section \ref{conn:sec:systemmodel}), we assume that an attacker is able to compromise a number of $a$ nodes.
We require that information exchange in the network is still possible even under this condition.
%Thus, the connectivity graph $D(V,E)$ of the CPS must still be connected.
Hence, to tolerate those $a$ compromised nodes, we need an $r$-resilient network with $r \geq a$.
This is fulfilled when the connectivity $\kappa(D)$ is greater than the necessary resilience, i.e., $\kappa(D) > r$.
Since each compromised node can disconnect at most one of the $\kappa(D)$ node-disjoint paths (cf.\ Section \ref{conn:sec:connectivity}), there is still at least one path remaining.
The correlation between the graph connectivity, the resilience and the number of attackers is summarized in Equation \ref{con:eq:resilience}.
\begin{equation}\label{con:eq:resilience}
  \kappa(D) > r \geq a
\end{equation}
From this equation, one can determine (1) the resilience of a given network as $r=\kappa(D)-1$ and (2) the required connectivity of a network for a specific $a$ as $\kappa(D) > a$.

%As consequence the maximum resilience $r_{max}$ of a CPS equals $\kappa(D)-1$.

%We require our distributed CPS to be $r$-resilient against attacks and failures
%We define $r$ as the \emph{resilience level} of the CPS communication network.
%
%We assume our system to be $r$-resilient towards attacks and failures. This means that up to $r$ arbitrary system components can fail and the system is still able to fulfill its task.
%
%
%Referenz zu \cite{heck2016multi} verarbeiten (wurde aus system model entfernt).
%TODO

\section{Evaluation}
\label{conn:sec:evaluation}

In this section, we first describe our simulation environment, i.e., the tools used to determine the network connectivity. After that, we present our evaluation methodology and the simulation scenarios. Finally, we present the achieved results and discuss them.

\subsection{Environment}
\label{conn:subsec:environment}

% Simulator Environment
% - PeerSim
% - HIPR
% - DIMACS not needed?
% - evtl HW

For our simulations, we use the network simulation software PeerSim \cite{montresor2009peersim}.
It is implemented with the Java programming language and includes an event protocol class for event driven simulations. We added Kademlia as an instance of this ``EDProtocol''.
Additionally we wrote software components to provide functionality for creating network churn (addition and removal of nodes) as well as requesting data objects and disseminating information into the network.

For the graph transformation, we implemented Even's algorithm in Java.
To calculate the maximum flow between a pair of vertices, we use the software ``HIPR'' \cite{goldberg_hipr}. It is a C implementation of the hi-level variant of the push-relabel algorithm presented in \cite{cherkassky1995implementing}.
In its original form, HIPR only calculates the maximum flow for one vertex pair.
Therefore, we modified it to support calculations with multiple vertex pairs per program invocation.
As adjacent vertex pairs do not influence the graph connectivity in our context (cf.~\ref{conn:subsec:vertex_connectivity_pair}), we also added program logic to detect such pairs.
We further wrote multiple software tools and scripts for both generation of maximum flow computing tasks, and validation and aggregation of the output created from these tasks.

We ran our simulations on a dual socket system with Intel Xeon E5-2690 CPUs (2.6 GHz), each with 14 cores plus hyper-threading. For the maximum flow computations, we used a Linux cluster provided by our University. We distributed the computations to 24 cluster nodes, each providing two 16 core AMD Opteron 6276 CPUs (2.3 GHz) with hyper-threading.

\subsection{Methodology}
\label{conn:subsec:methodology}

% Methodology (Vorgehen)
%- Wie haben wir den max flow berechnet
%- Time stamps + max flow calculation
%- partial max flow calculation

To calculate the graph connectivity over time, we persist the connectivity graph of a network at pre-defined time stamps in a simulation.
For that purpose, we interrupt the simulation and save the current contents of the routing tables of all network nodes to disk into a snapshot file.
We use this snapshot file to transform the connectivity graph with Even's algorithm.
Next, we convert the transformed graph to the supported input format of HIPR (i.e., DIMACS \cite{dimacs}) to calculate the maximum flow.

The push-relabel algorithm used for the maximum flow computation for a single vertex pair in HIPR has a worst case time complexity of $\mathcal{O}(n^2 \sqrt{m})$, where $n$ is the number of vertices and $m$ the number of edges in the processed graph \cite{cherkassky1995implementing}.
Since the transformed graph $D'(V',E')$ contains $2n$ nodes and $n+m$ edges, the complexity of calculating the maximum flow of a single vertex pair in $D'$ is $\mathcal{O}(n^2 \sqrt{n+m})$.
To calculate the graph connectivity $\kappa(D')$, we need to apply the above calculation on the transformed graph from all outgoing vertices to all incoming vertices, i.e., $n(n-1)$ times.
Thus, the overall time complexity for calculating $\kappa(D')$ is $\mathcal{O}(n^4 \sqrt{n+m})$. This complexity makes the maximum flow computation very expensive. For instance, the full maximum flow computation for a transformed connectivity graph with $2500$ vertices takes about 250 hours on a single CPU core.

The nodes in Kademlia attempt to add each other to their respective routing tables. This would result in an undirected connectivity graph.
However, due to size restrictions of the buckets in the routing table and race conditions, these attempts are not always successful.
Hence, there is no guarantee for the connectivity graph being undirected.
Nevertheless, our analysis of simulation runs shows that the connectivity graphs come very close to being undirected.
This allows us to reduce the amount of maximum flow computations from $n(n-1)$ to $c \cdot n(n-1)$, $0 < c \leq 1$.
We achieve this reduction by only using a percentage $c \cdot n$ of outgoing vertices for the maximum flow calculation.
%The reduction is exclusively done by not using all $n$ outgoing vertices as source vertices for the maximum flow computations, but only $c \cdot n$ of them.
Since the outgoing degree $d_{\mathit{out},v}$ of a vertex $v$ is an upper limit for the outgoing flow, we select those $c \cdot n$ outgoing vertices with the smallest $d_{\mathit{out}}$.
As we calculate the maximum flow from only a percentage $c \cdot n$ of outgoing vertices to all $n-1$ incoming vertices, also the limiting incoming degree $d_{\mathit{in}}$ is still considered.
We verified this with 20 randomly selected connectivity graphs, for which we performed a full analysis, i.e., calculated the maximum flow for all $n(n-1)$ vertex pairs.
In all cases, $c=0.02$ (2\%) was sufficient to determine the minimum of the maximum flows, i.e., the graphs vertex connectivity.

\subsection{Scenarios}
\label{conn:subsec:scenarios}

%%%%%%%%%%%%%%%%%%%%%%%%%%%%%%%%%%%%%%%%%%%%%%%%%
%An example for a CPS is a Smart Camera Network (SCN). Smart cameras are video cameras with a built-in computation unit that can be utilized for various tasks, e.g., image processing, object localization or object tracking.
%With their integrated wired or wireless communication devices, smart cameras are able to communicate with each other and can assess scenarios, which go beyond the scope of a single camera.
%Today, the most common scenario for a SCN are the detection of intruders in restricted areas or the surveillance of high risk areas. On detection, the system could, e.g., notify security personnel or activate stronger access restrictions.
%Another example for a CPS is a distributed network intrusion detection system (IDS) for securing cooperate networks with several branches.
%At each branch, multiple nodes observe and classify all passing traffic as \emph{suspicious} or \emph{normal} to detect and handle attacks.
%The nodes exchange their observed information regularly to increase the chance of detecting distributed attacks.
%Those might not be detected by individual nodes, but with the aggregated information the detection possibility increases and allows for faster and more efficient adaptation by each node.
%%%%%%%%%%%%%%%%%%%%%%%%%%%%%%%%%%%%%%%%%%%%%%%%%%%%%%%%%%%%

In a two page short paper~\cite{heck2016evaluating}, we briefly presented some shorter simulations done with an earlier version of our simulator, varying a single Kademlia parameter, namely $k$.
Based on these, we have made several improvements.
Previously, we investigated three different mechanisms for node bootstrapping, which turned out to show no significant differences with regard to connectivity. Therefore, we now apply only one bootstrap mechanism, in which the bootstrap nodes are completely random, and any node can be affected by churn. Also, to bring the simulations closer to a real world scenario, actions affecting the network structure, e.g., lookup procedures and node removals, are done at random points in time within a predetermined time frame.
As a result, the \emph{initial bootstrap procedure} to create the network is done randomly in terms of time and bootstrap node selection. A new node joins the network at a random point in the simulated time that is evenly distributed between 0 and 30 minutes. The bootstrap node is randomly chosen from the already joined nodes.

Beyond that we extended the number of varied Kademlia parameters in our simulations from one to four and also added scenarios with communication channels affected by message loss.
To determine how different environments and protocol parameters influence the connectivity of the network, we devised a total eight dimensions for the simulations, i.e., network size, network churn, network traffic, message loss, the Kademlia bucket size $k$, the parallelism factor $\alpha$, the bit-length $b$, and the staleness limit $s$.

\subsubsection*{Network Size}

We consider two different scenarios for the network size, i.e., a small network with 250 nodes and a large one with 2500 nodes.
Our choice for these network sizes is based on the CPS examples introduced in Section \ref{conn:sec:introduction}.
For the smart camera scenario, a large number of smart cameras may be necessary for reliably observing and controlling an industrial complex. Thus, we simulate it with $250$ nodes.
In contrast, a distributed IDS can be used for securing corporate networks spanning several branches.
Such networks usually comprise several hundreds to thousands of nodes. Exemplarily, we choose $2500$ nodes for this scenario.

\subsubsection*{Network Churn}

We consider three different churn scenarios.
In the scenario (0/1), we remove a single node from the network in every minute of simulated time and add no nodes. In the scenario (1/1), we add one node and remove one node every minute. Similarly, in the scenario (10/10), we remove ten nodes and add ten nodes per minute. The add/remove actions happen at random points in time within each minute range. We chose these high churn rates to to get a clear indication of effects related to churn in our simulations.

\subsubsection*{Network Traffic}

We distinguish two different scenarios with respect to data traffic, i.e., with and without data traffic.
In the scenario with data traffic, all nodes regularly look up data objects and disseminate them.
For this, each node performs $10$ lookup procedures an $1$ dissemination procedure per minute during the whole simulation.
The procedures happen at random points in time within each minute range.
In the scenario without data traffic, the nodes do not lookup data objects or disseminate them.
However, for maintenance purposes Kademlia requires each node to perform a so-called ``bucket-refresh'' every 60 minutes.
For this, a node randomly generates an \emph{id} from the \emph{id} range of each $k$-bucket and performs lookup procedures for these \emph{ids}.
This way, it can learn about previously unknown contacts and stale contacts in its routing table.
Hence, even in the scenario without data traffic, there is some basic maintenance traffic.

\subsubsection*{Message Loss}

Since two-way communication in the form of request/response is the most used communication type in Kademlia,
we tailor our message loss $l$ towards it. We apply four different message loss scenarios with different probabilities for a two-way communication to fail.
Those probabilities apply to any communication taking place between nodes.
The first scenario, $none$, has no loss at all, all messages reach their destination. Unless marked otherwise, this is the default case.
The three other scenarios are $low$, $medium$ and $high$. Table \ref{conn:table:lossscenarios} shows the loss probabilities for one-way and two-way communication for all four scenarios.

\begin{table}[ht]
  \centering
  %\bgroup
  \def\arraystretch{1.05}%  1 is the default, change whatever you need
  \hspace*{0.7cm}
  \begin{tabular}{|l|c|c|}
  \hline
   \multicolumn{1}{|c|}{\textbf{Loss }$\mathbf{l}$} & $\mathbf{P_{loss}}$\textbf{(1-way)} & $\mathbf{P_{loss}}$\textbf{(2-way)} \\ \hline
   $none$ & 0.0\% & 0\% \\
   $low$ & 2.5\% & 5\% \\
   $medium$ & 13.4\% & 25\% \\
   $high$ & 29.3\% & 50\% \\ \hline
\end{tabular}
  \newline
  \caption{Message loss scenarios with loss probabilities for one way and two way communication.}\label{conn:table:lossscenarios}
\end{table}

\subsubsection*{Kademlia Bucket Size}

In Kademlia, the bucket size $k$ is directly responsible for the number of contacts a node can keep in its routing table.
To determine its effect on the network connectivity, we use four different values for $k$, i.e., $k\in\{5,10,20,30\}$.

\subsubsection*{Kademlia Request Parallelism}

The request parallelism $\alpha$ determines how many queries are made in parallel when locating a node or data object.
We use the values $3$ and $5$ for $\alpha$.

\subsubsection*{Kademlia Staleness Limit}

The staleness limit $s$ is the number of unsuccessful communication attempts in a row that lead to the removal of a contact from the routing table, assuming it has left the network.
We use the values $1$ and $5$ for $s$. In simulations with churn, which are not specifically meant for evaluating $s$ and have the loss scenario $none$, we set $s=1$. This allows quick reaction to nodes leaving the network and, therefore, provides a clearer picture on the influence of churn.

\subsubsection*{Kademlia Bit-length}

The bit-length $b$ is the size of the numerical identifier of a node or data object in bits.
We use the values $160$ and $80$ for $b$.

In summary, we have eight dimensions with several scenarios for each of them, i.e., $1536$ combinations. We simulated a majority of these combinations to determine how the dimension and the connectivity correlate and present our results in the next section.

\subsection{Simulation Phases}

In all simulations the network is fully setup after 30 minutes (setup phase), as described in detail above.
From minute 30 to minute 120 (stabilization phase), we allow the network to stabilize. These 90 minutes guarantee that for scenarios without data traffic each node performs a bucket refresh. After that, starting at minute 120, we apply a churn scenario, if the simulation requires churn (churn phase).

\subsection{Results for Traffic, Churn, and Bucket Size $k$}

% for 250 (SCN)
% for 2500 (IDS)

In this section, we present the simulations and measurement results for different network scenarios.
The first four simulations focus on the effect of traffic, while the remaining simulations focus on the effect of churn.
In each graph, we present the simulations for all four bucket sizes.

\subsubsection{Without data traffic}

\begin{figure}[ht]
  \centering
  \includegraphics[width=0.98\columnwidth]{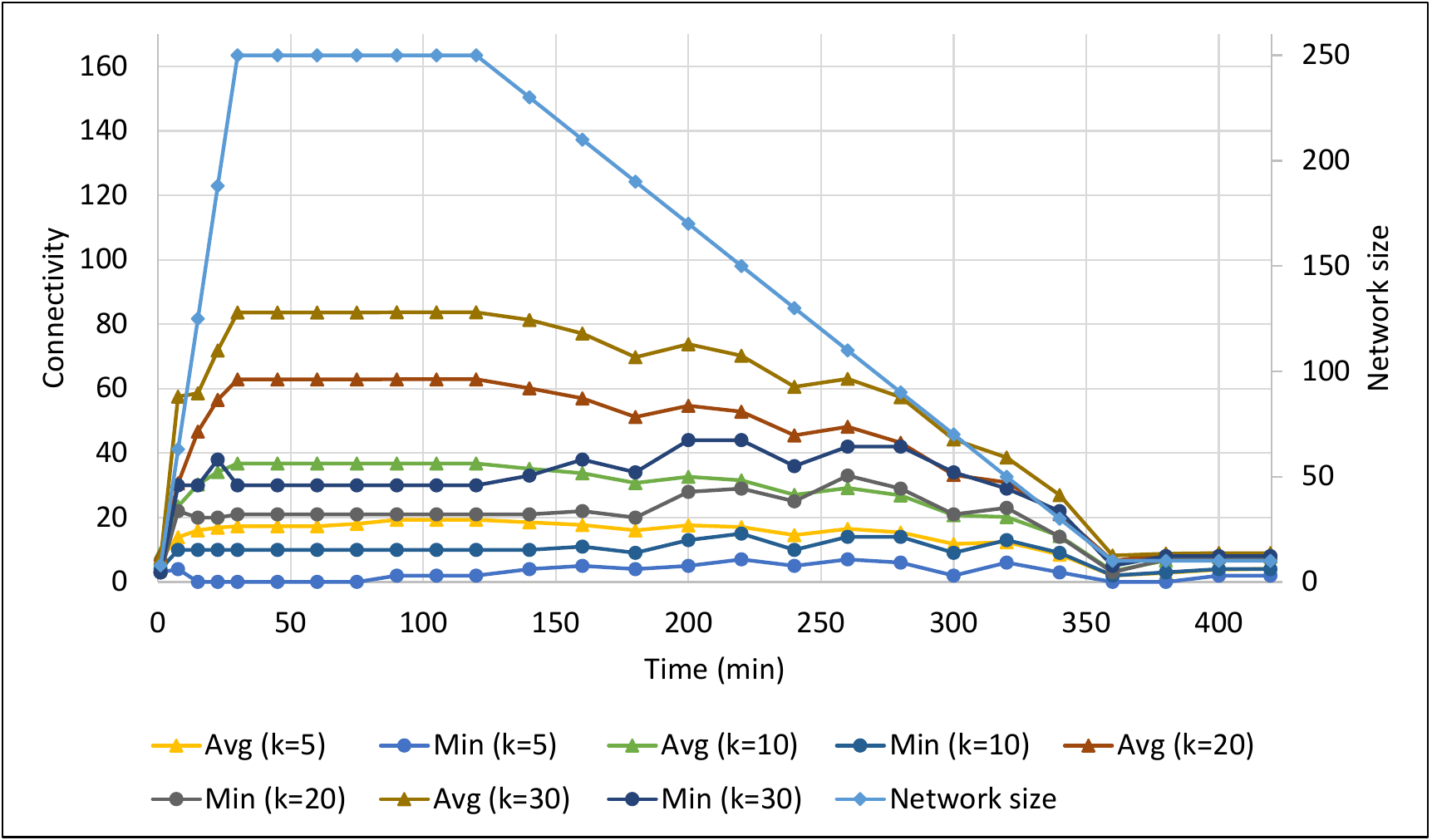}\\
  \caption{Simulation\,A, Size 250, churn 0/1, without data traffic}\label{conn:figure:200-203-crop}
\end{figure}

\begin{figure}[ht]
  \centering
  \includegraphics[width=0.98\columnwidth]{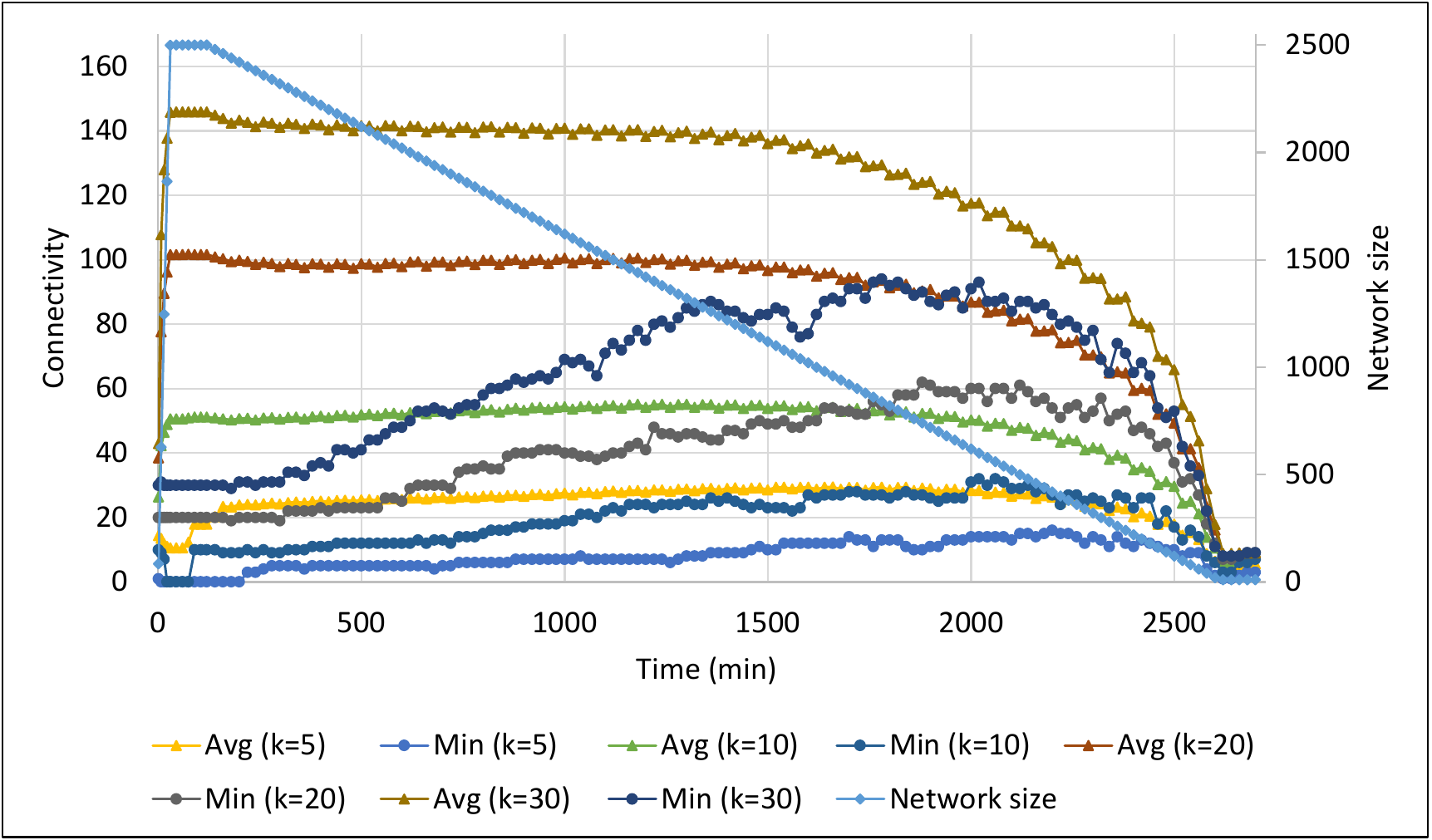}\\
  \caption{Simulation\,B, Size 2500, churn 0/1, without data traffic}\label{conn:figure:208-211-crop}
\end{figure}

In Simulations A \& B, no data traffic is present. The churn scenario is 0/1.
We present the simulation for the network size 250 in Figure~\ref{conn:figure:200-203-crop}
and the simulation for the network size 2500 in Figure~\ref{conn:figure:208-211-crop}.

After the setup phase, the connectivity for $k\in\{20,30\}$ is at roughly $k$ for both network sizes. For $k=10$, this is also true for the small network, whereas the connectivity is zero in the large network. For $k=5$, the connectivity is zero in both networks. For the smaller $k$ values, the setup appears to be more problematic the bigger the network.
Further investigation showed, that this is caused by a single digit number of disconnected nodes.
While those nodes do not have significantly less entries in their own routing table than others, they themselves only appear in the routing tables of less than $k$ other nodes or none at all.
This issue is resolved during the stabilization phase for $k=10$, so that the connectivity is roughly $k$ for $k\in\{10,20,30\}$.
In the churn phase, the minimum connectivity first increases overall for all $k$ values.
This effect also applies to $k=5$, so that now the minimum connectivity for the smallest $k$ value rises to $k$ and above.
With continuing churn and decreasing network size, the minimum connectivity drops again.

It appears that the network state after stabilization is not ideal from a connectivity point of view.
The leaving nodes enable the network to reconfigure (freed up entries in the $k$-buckets) towards a higher connectivity.
This continues until the network size becomes too small to sustain this behavior.
%Towards the end of the simulation, with 10 nodes left in the network, the network is fully connected for each bucket size except the smallest one.

%\subsubsection*{Simulations 3 \& 4 (Figures~\ref{conn:figure:204-207-crop} \& \ref{conn:figure:212-215-crop})}

\subsubsection{With data traffic}

\begin{figure}[ht]
  \centering
  \includegraphics[width=0.98\columnwidth]{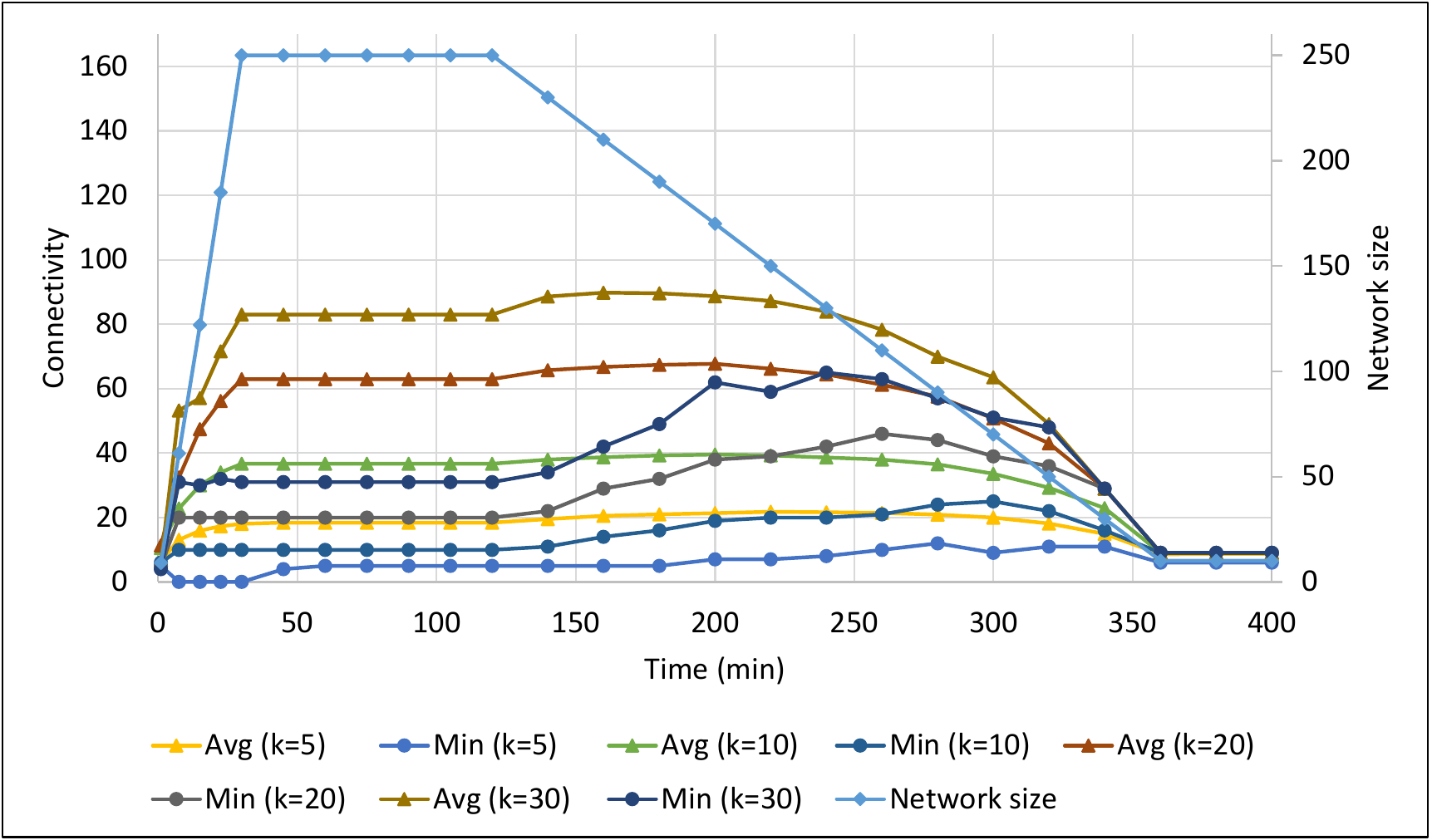}\\
  \caption{Simulation\,C, Size 250, churn 0/1, with data traffic}\label{conn:figure:204-207-crop}
\end{figure}

\begin{figure}[ht]
  \centering
  \includegraphics[width=0.98\columnwidth]{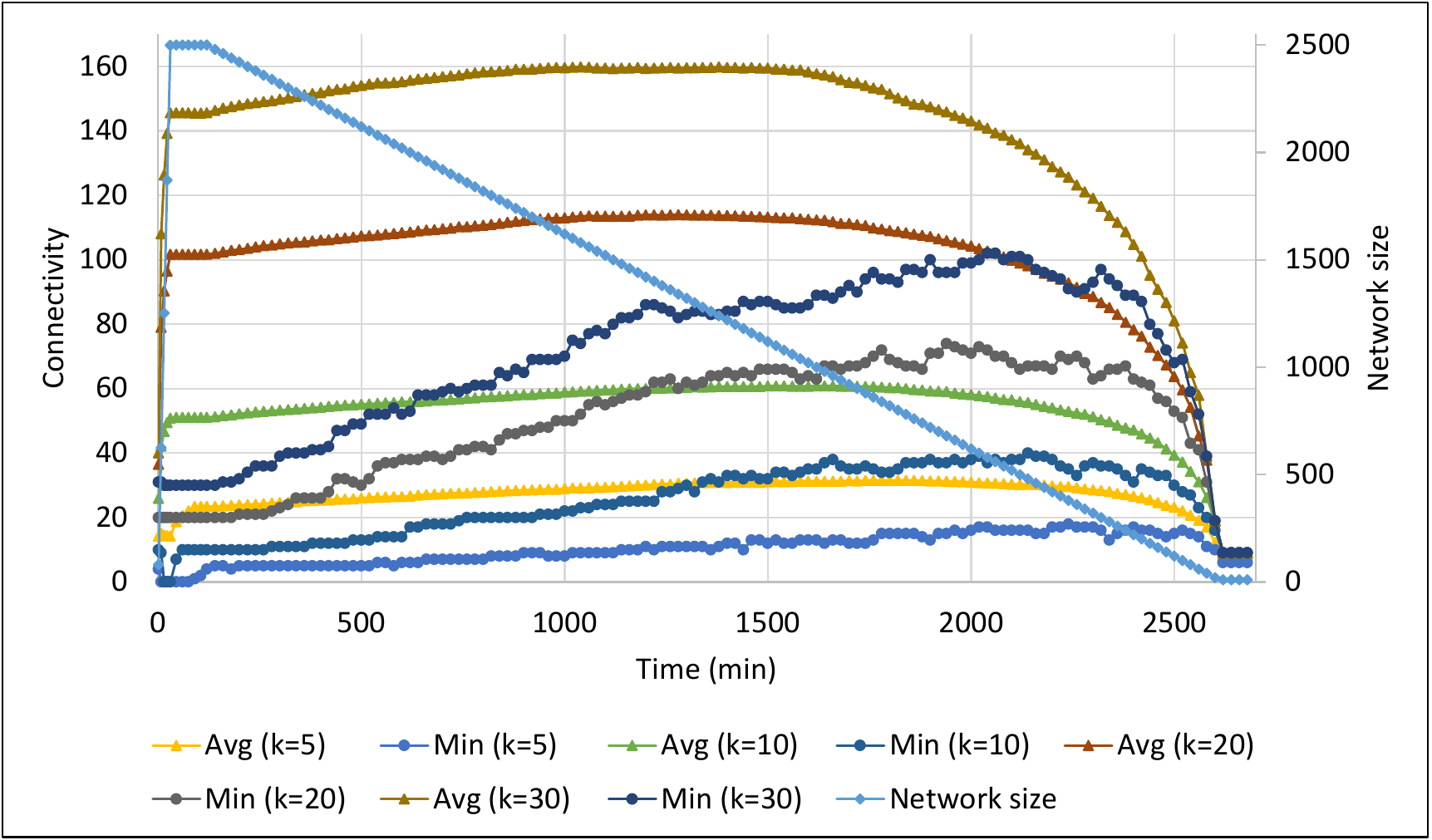}\\
  \caption{Simulation\,D, Size 2500, churn 0/1, with data traffic}\label{conn:figure:212-215-crop}
\end{figure}

In Simulations C \& D, data traffic is present. The churn scenario is 0/1.
We present the simulation for the network size 250 in Figure~\ref{conn:figure:204-207-crop}
and the simulation for the network size 2500 in Figure~\ref{conn:figure:212-215-crop}.

The setup phase is similar to that in Simulations A \& B. At its end, the connectivity for $k\in\{20,30\}$ is at roughly $k$ for both network sizes. For $k=10$, this is also true for the small network, whereas the connectivity is zero in the large network. For $k=5$, the connectivity is zero for both network sizes. For the smaller $k$ values, the setup again appears to be more problematic the bigger the network. The cause are, as before, a single digit number of disconnected nodes, which do not appear in the routing tables of other nodes.
This issue is resolved during the stabilization phase for all four $k$ values, so that the connectivity is roughly $k$.
In the churn phase, the minimum connectivity first increases overall for all $k$ values.
With continuing churn and decreasing network size, the minimum connectivity drops again.

Compared to Simulations A \& B, the observed effects are similar, but their timing and strength are different.
Connectivity values of $k$ or above are reached earlier in the simulations.
The increase in minimum connectivity with churn is much more pronounced and its maximum values are also greater.
Towards the end of the simulation, with 10 nodes left in the network, the network is now fully connected for each bucket size except the smallest one.
As one would expect, the data traffic results in an overall improved connectivity.

\subsubsection{With 1/1 churn}

\begin{figure}[ht]
  \centering
  \includegraphics[width=0.98\columnwidth]{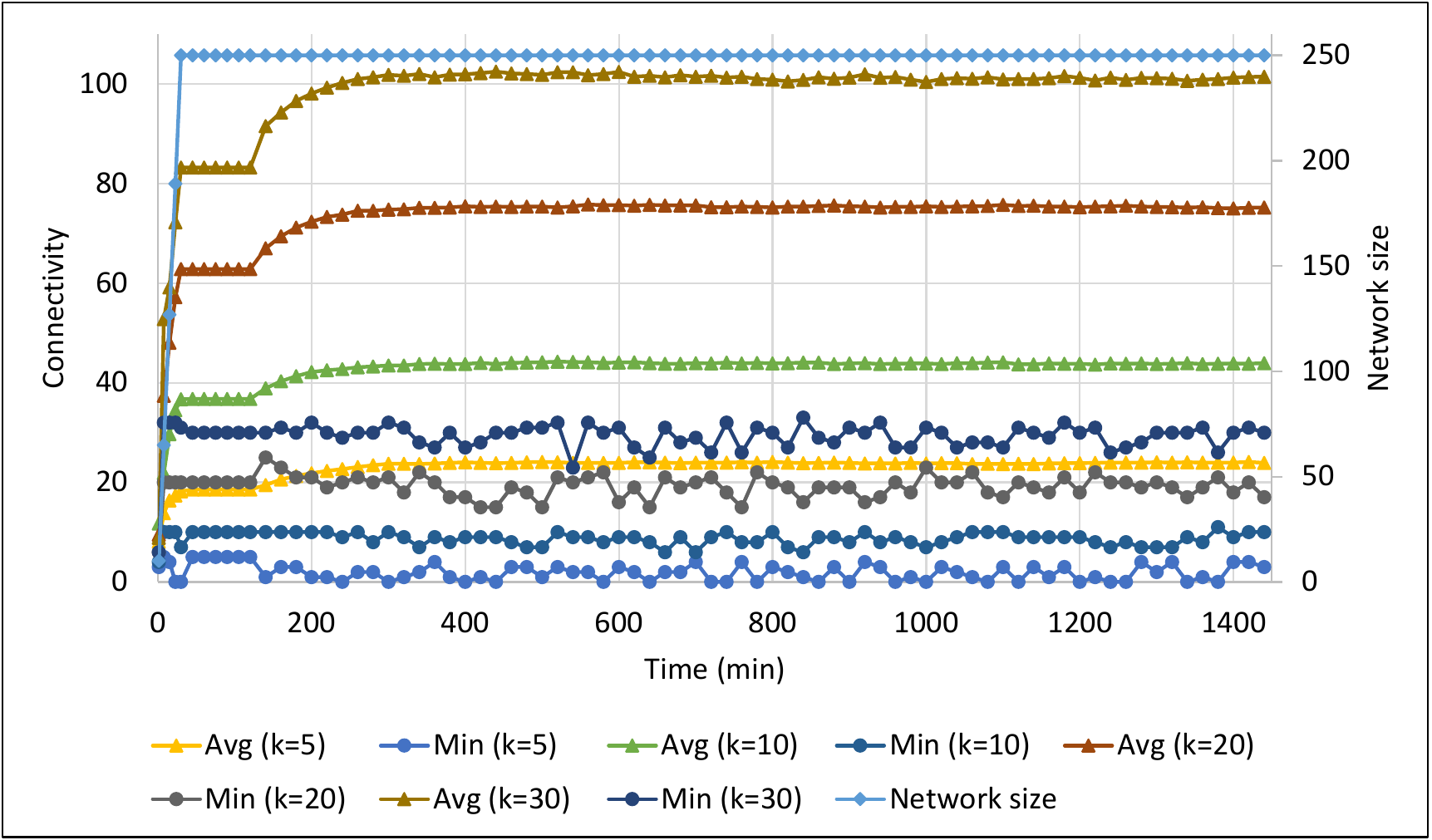}\\
  \caption{Simulation\,E, Size 250, churn 1/1, with data traffic}\label{conn:figure:216-219-crop}
\end{figure}

\begin{figure}[ht]
  \centering
  \includegraphics[width=0.98\columnwidth]{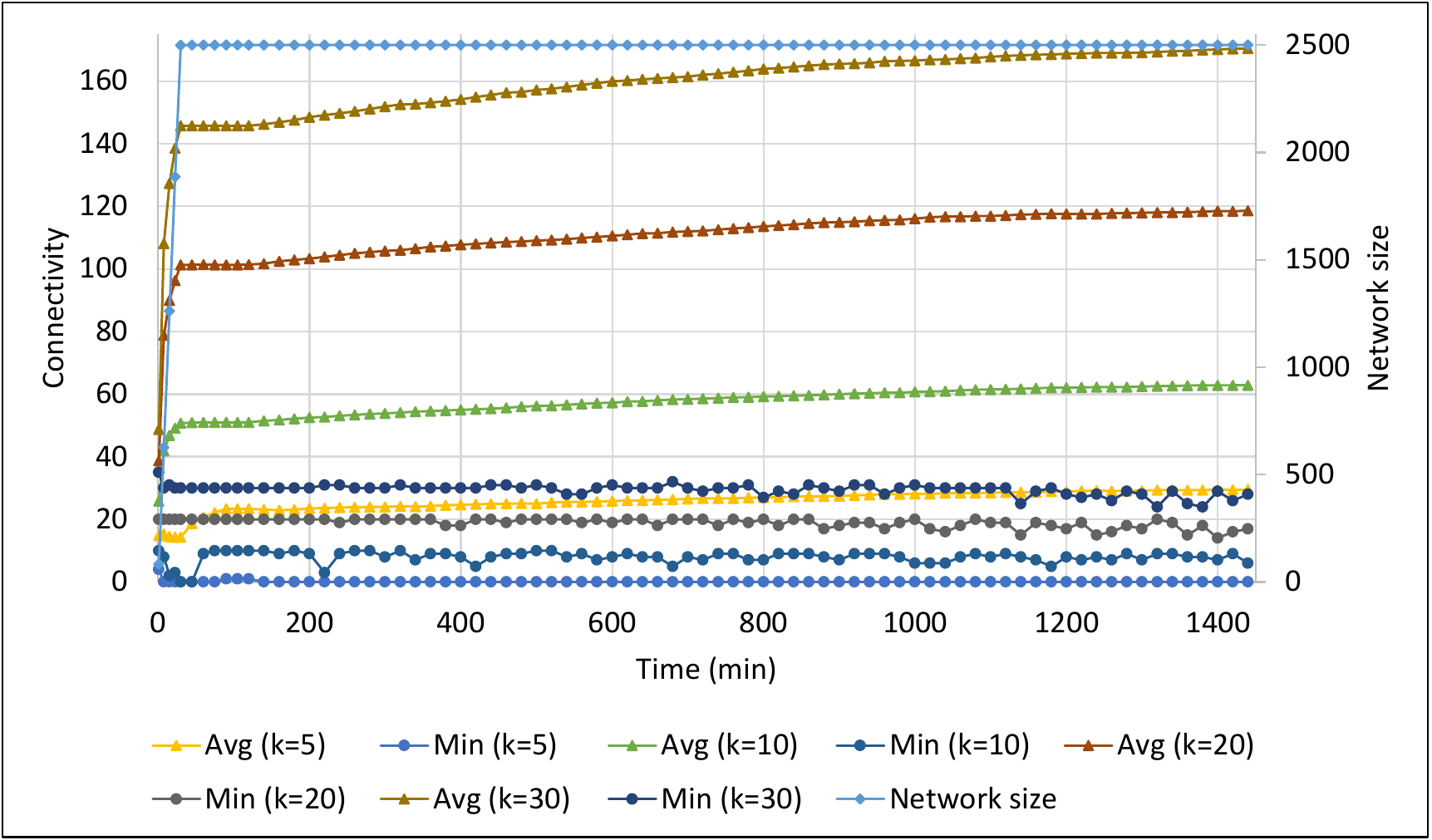}\\
  \caption{Simulation\,F, Size 2500, churn 1/1, with data traffic}\label{conn:figure:220-223-crop}
\end{figure}

In Simulations E \& F, data traffic is present. The churn scenario is 1/1.
We present the simulation for the network size 250 in Figure~\ref{conn:figure:216-219-crop}
and the simulation for the network size 2500 in Figure~\ref{conn:figure:220-223-crop}.

As this is a simulation with data traffic, the setup phase and stabilization phase are similar to those in Simulations C \& D.
An exception here is the large network with $k=5$. Its minimum connectivity does not quite reach $k$ at the end of the stabilization phase, but is none the less greater than zero.

Whereas, similar to the 0/1 churn, the average connectivity benefits from the churn phase, the minimum connectivity does not.
For the greater values of $k$ the minimum connectivity oscillates around $k$, for smaller values it drops significantly, even down to 0.
This effect is more pronounced in the larger network, where for $k=5$ the minimum connectivity is 0 throughout almost the whole churn phase.

\subsubsection{With 10/10 churn}

\begin{figure}[ht]
  \centering
  \includegraphics[width=0.98\columnwidth]{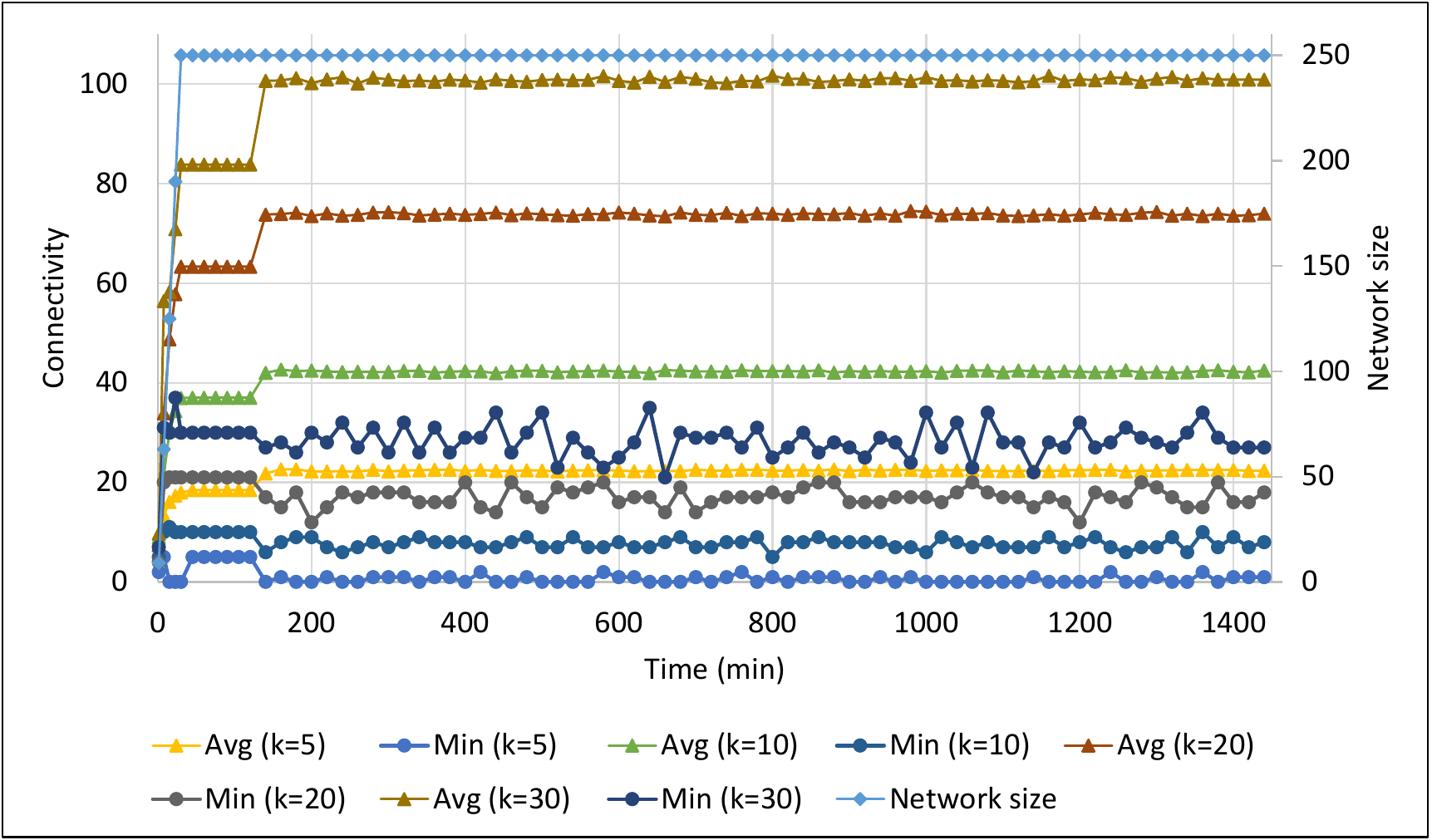}\\
  \caption{Simulation\,G, Size 250, churn 10/10, with data traffic}\label{conn:figure:224-227-crop}
\end{figure}

\begin{figure}[ht]
  \centering
  \includegraphics[width=0.98\columnwidth]{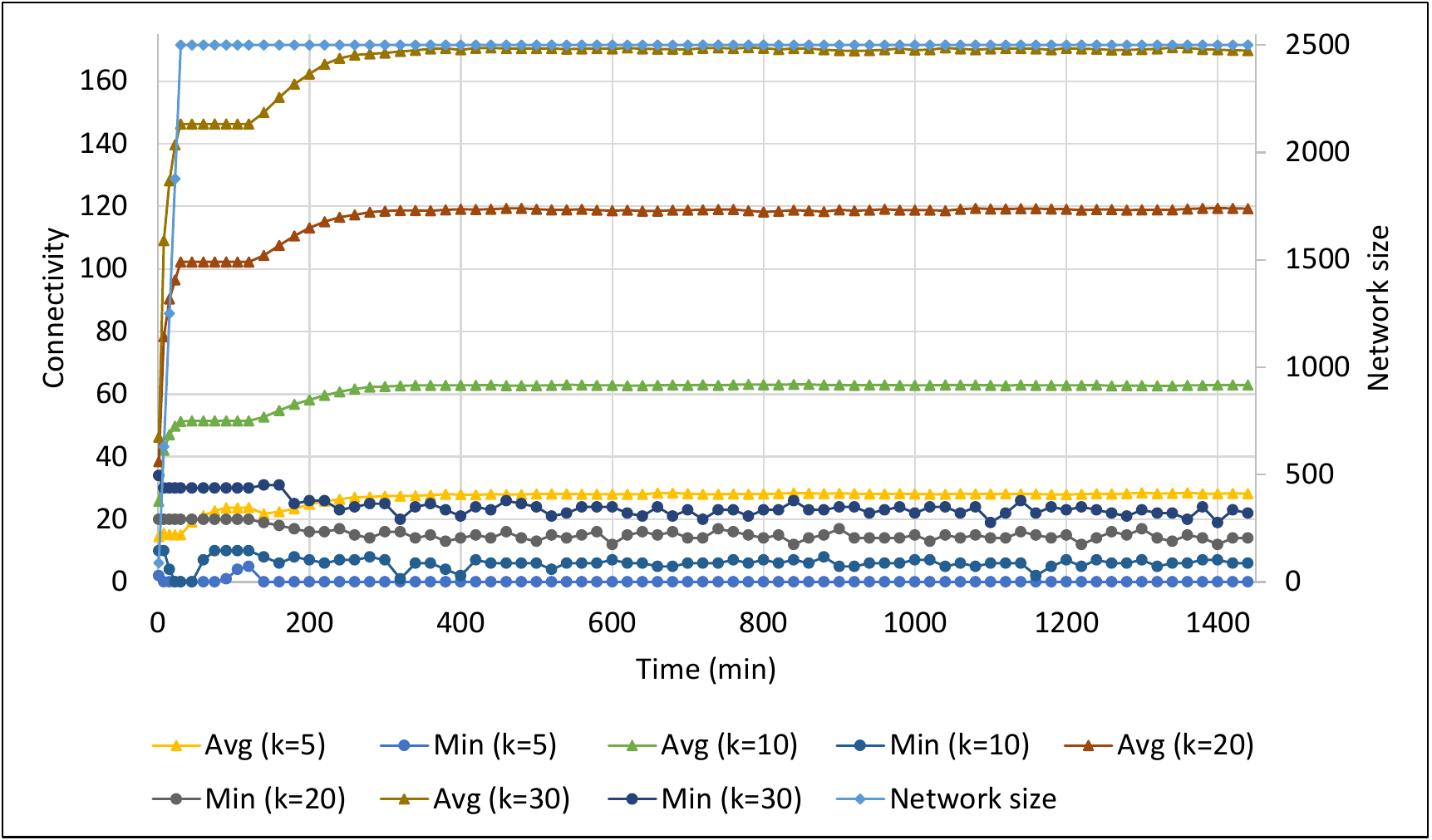}\\
  \caption{Simulation\,H, Size 2500, churn 10/10, with data traffic}\label{conn:figure:228-231-crop}
\end{figure}

In Simulations G \& H, data traffic is present. The churn scenario is 10/10.
We present the simulation for the network size 250 in Figure~\ref{conn:figure:224-227-crop}
and the simulation for the network size 2500 in Figure~\ref{conn:figure:228-231-crop}.

The setup phase and stabilization phase are basically identical to those in Simulations C \& D.

With the increased churn, the average connectivity reaches the same levels as with simulations E \& F, but rises much faster as soon as the churn sets in.
For the minimum connectivity the differences are more significant. Where the absolute values allow it, the oscillation increases.
The overall level drops for all $k$ values, so that e.g.~the minimum connectivity for $k=5$ is now almost always at 0 also for the small network.

\subsubsection{Relative Variance for Churn}
A numerical comparison of the 1/1 and 10/10 churn scenarios is given in Table \ref{conn:table:variance}.
It shows the means and the Relative Variance (RV), i.e., Variance/Mean, of the minimum connectivity during the churn phase for the simulations E~to~H for all four $k$ values.
From the graphs our impression was that increased churn leads to increased variance relative to the connectivity.
%To quantify this, we calculated the variance. However, this did not show the expected result.
%The reason for this is that the mean of the minimum connectivity often drops significantly with the increased churn, severely limiting the dynamic.
We, therefore, calculated the RV to express the effects of increased churn.
As the RV values in Table \ref{conn:table:variance} show, the increase in churn from 1/1 to 10/10 leads to an increased RV in all simulations. The exception is the network size 2500 with $k=5$, where the minimum connectivity is zero throughout the whole churn phase for both churn scenarios.

\begin{table}[ht]
  \centering
  %\bgroup
  \def\arraystretch{1.05}%  1 is the default, change whatever you need
  \hspace*{0.7cm}
  \begin{tabular}{|c|c|c|r|c|}
  \hline
   \textbf{Size} & $\mathbf{k}$ & \textbf{Churn} & \textbf{Mean} & \textbf{RV } \\ \hline
   \multirow{8}{*}{250} & \multirow{2}{*}{5} & $1/1$ & 3.49 & 0.63 \\ %\cline{3-5}
   &  & $10/10$ & 1.93 & 0.75 \\ \cline{2-5}
   & \multirow{2}{*}{10} & $1/1$ & 10.12 & 0.17 \\ %\cline{3-5}
   &  & $10/10$ & 9.22 & 0.23 \\ \cline{2-5}
   & \multirow{2}{*}{20} & $1/1$ & 22.22 & 0.36 \\ %\cline{3-5}
   &  & $10/10$ & 20.53 & 0.39 \\ \cline{2-5}
   & \multirow{2}{*}{30} & $1/1$ & 32.84 & 0.34 \\ %\cline{3-5}
   &  & $10/10$ & 32.78 & 0.62 \\ \cline{1-5}

    \multirow{8}{*}{2500} & \multirow{2}{*}{5} & $1/1$ & 0.00 & 0.00 \\ %\cline{3-5}
   &  & $10/10$ & 0.00 & 0.00 \\ \cline{2-5}
   & \multirow{2}{*}{10} & $1/1$ & 9.30 & 0.13 \\ %\cline{3-5}
   &  & $10/10$ & 7.38 & 0.21 \\ \cline{2-5}
   & \multirow{2}{*}{20} & $1/1$ & 22.06 & 0.07 \\ %\cline{3-5}
   &  & $10/10$ & 16.62 & 0.16 \\ \cline{2-5}
   & \multirow{2}{*}{30} & $1/1$ & 31.35 & 0.10 \\ %\cline{3-5}
   &  & $10/10$ & 25.73 & 0.24 \\ \cline{1-5}

\end{tabular}
  \newline
  \caption{Simulations E to H: Means and Relative Variance (RV)}\label{conn:table:variance}
\end{table}

\subsection{Results for Request Parallelism $\alpha$}

\begin{figure}[ht]
  \centering
  \subfloat[Network size 250 (small)\label{conn:figure:avgmins250}]{\includegraphics[width=0.98\columnwidth]{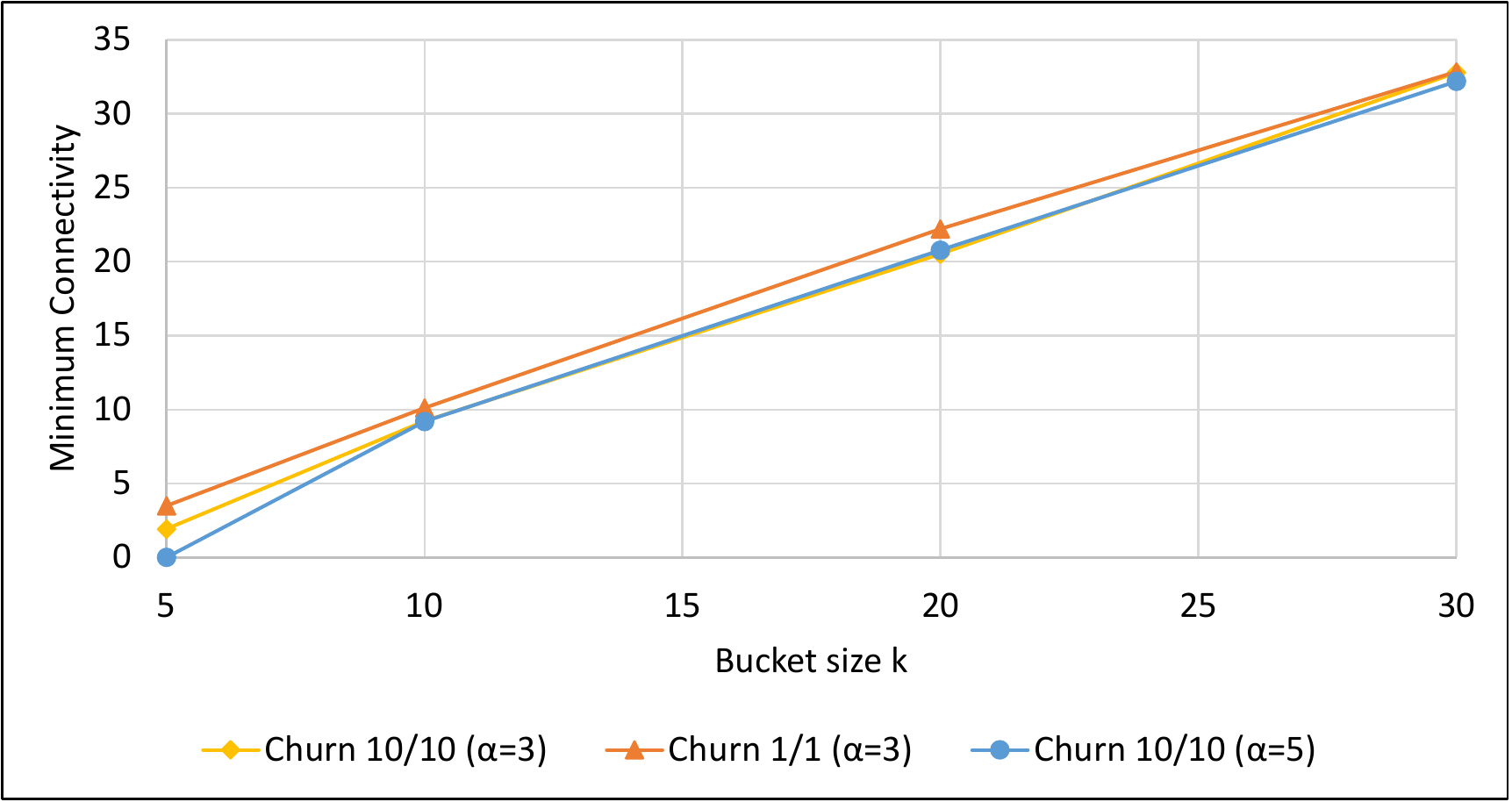}}\qquad
  \subfloat[Network size 2500 (large)\label{conn:figure:avgmins2500}]{\includegraphics[width=0.98\columnwidth]{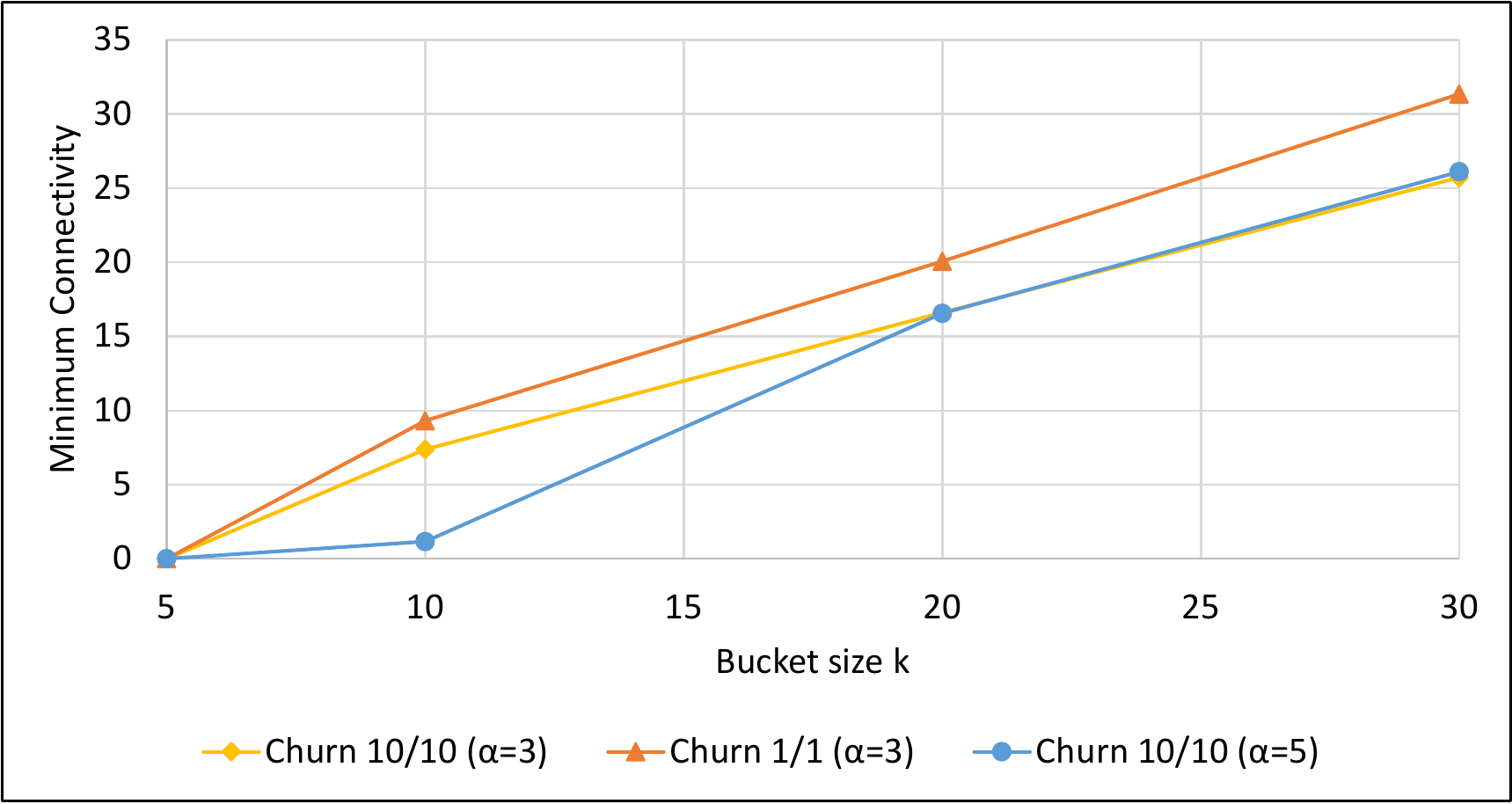}}\qquad
 % \includegraphics[width=\columnwidth]{images/avg_mins_250-crop}\\
  \caption{Means of the minimum connectivity during churn}\label{conn:figure:avgmins}

 % \includegraphics[width=\columnwidth]{images/avg_mins_2500-crop}\\
 % \caption{Simulation\,H, Size 2500, churn 10/10, with data traffic}\label{conn:figure:228-231-crop}

\end{figure}

Figures \ref{conn:figure:avgmins250} and \ref{conn:figure:avgmins2500} show the means of the minimum connectivity during churn for all four $k$ values for simulations E~to~H (see Table \ref{conn:table:variance}) and for additional simulations. For simulations E~to~H the request parallelism $\alpha$ equals 3. The additional simulations have the same scenarios as G (small network, churn 10/10) and H (large network, churn 10/10), except now $\alpha$ equals 5.
The figures show the following: 1) The scenarios with churn 1/1 show a higher connectivity than those with churn 10/10. This is more prominent in the large network, than in the small one.
2) For $k=5$ the connectivity is zero for all scenarios of the large network and for churn 10/10 with $\alpha=5$ in the small network. Therefore, $k\geq10$ seems to be the minimum advised $k$ for a connected network. 3) The increase of $\alpha$ from 3 to 5 with churn 10/10 has a very negative impact on connectivity for the smaller $k$ values. The connectivity for $k=5$ is zero for both network sizes and almost zero for $k=10$ in the large network. This is possibly due to the fact that a node contacts more other nodes in parallel and, therefore, takes places in more routing tables. Those places are not available for joining nodes, so that for small $k$ a disconnected node is very likely.

\subsection{Results for Bit-length $b$}

In other simulations we used the the same scenarios as in C and D, except for the identifier size $b$, which changed from 160 to 80.
They showed no significant difference from C and D with regard to connectivity.

\subsection{Message Loss and Staleness Limit $s$}

In this section we present results from simulations with focus on message loss and the staleness limit $s$.
The following settings apply to all shown simulations: Data traffic is present, the bucket size $k$ is 20, and the bit-length $b$ is 160.
Since the observed effects are very similar for both network sizes, we present the results for the large network only.

\subsubsection{Staleness Limits without Message Loss}

\begin{figure}[ht]
  \centering
  \subfloat[Churn 1/1 \label{conn:figure:staleness_1_1}]{\includegraphics[width=0.98\columnwidth]{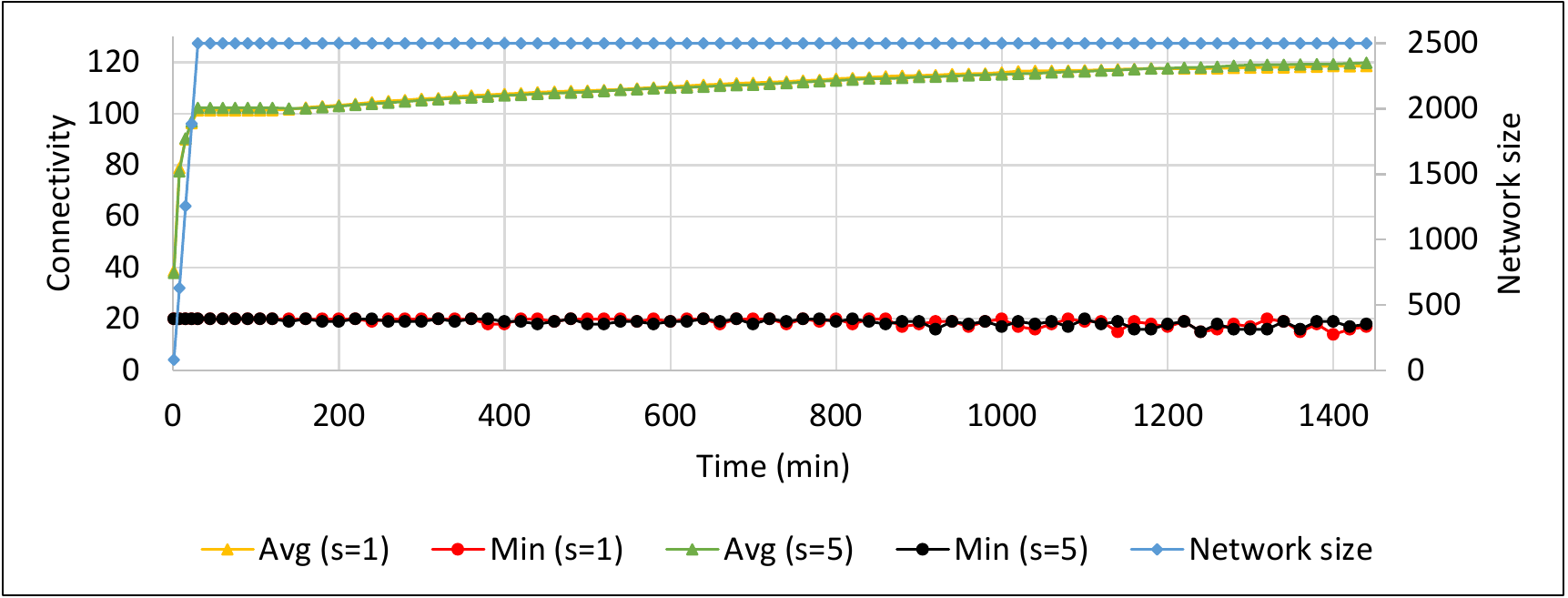}}\qquad
  \subfloat[Churn 10/10 \label{conn:figure:staleness_10_10}]{\includegraphics[width=0.98\columnwidth]{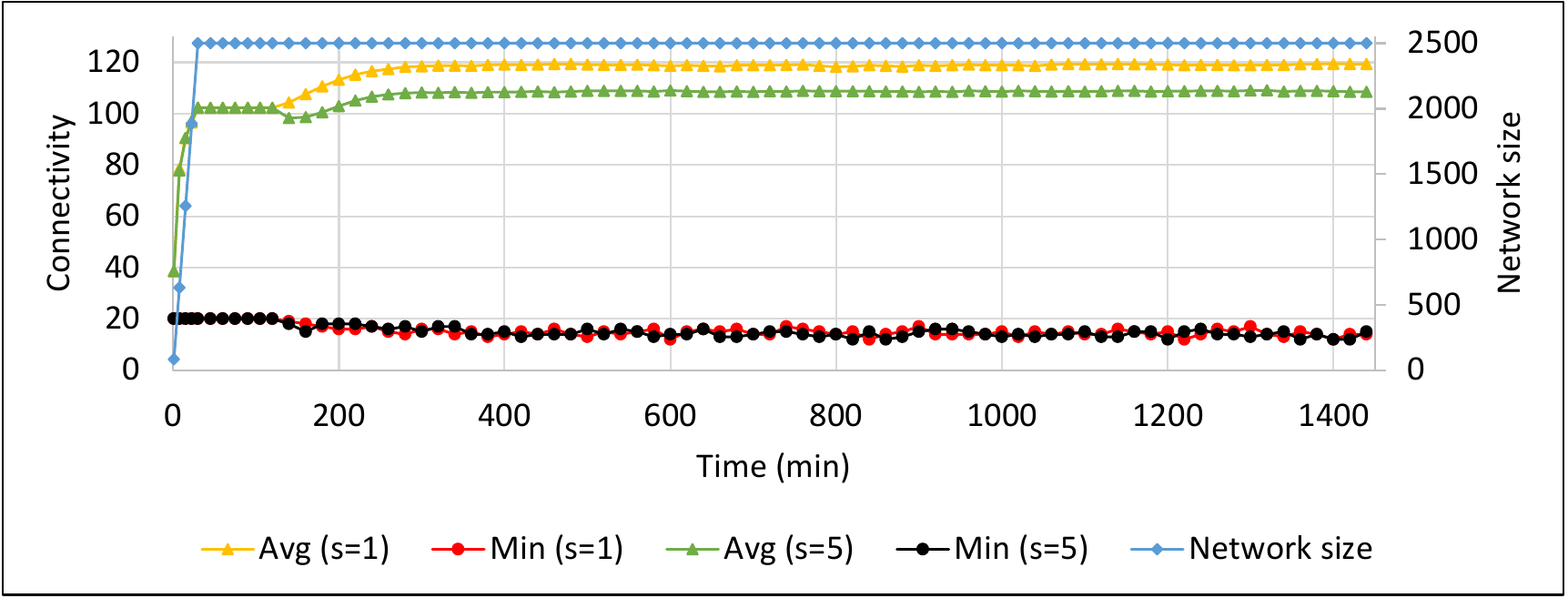}}\qquad
 % \includegraphics[width=\columnwidth]{images/avg_mins_250-crop}\\
  \caption{Simulation\,I, no message loss, $s\in\{1,5\}$}\label{conn:figure:staleness}
\end{figure}

Simulation I shows the effect of the two different staleness limits $s\in\{1,5\}$ in a network affected by churn and without message loss ($l$=$none$).
We present the simulation for churn 1/1 in Figure~\ref{conn:figure:staleness_1_1} and the simulation for churn 10/10 in Figure~\ref{conn:figure:staleness_10_10}.
With churn 1/1, there is no significant difference between the two staleness limits.
With churn 10/10, the average connectivity for $s$=$5$ drops below that of $s$=$1$, as soon as the churn phase begins. It remains that way for the remainder of the simulation.
Three effects are responsible for this:
1) With the stronger network churn more nodes become stale per minute, resulting in more stale routing table entries.
2) The stronger churn also results in more nodes joining per minute, each of them needing to become connected.
3) With the greater staleness limit it takes all nodes longer to detect and remove stale entries in their routing tables.
Since the routing table size is limited and each stale entry potentially keeps a new contact from entering the routing table, the average connectivity decreases.
Interestingly, the minimum connectivity is not affected.
It is the same for $s$=$1$ as for $s$=$5$.
At the moment we don't know the reason for this, but we will investigate this issue further.

\subsubsection{Staleness Limit with Message Loss}

The simulations J, K and L show the effect of the three message loss scenarios $l\in\{low,medium,high\}$ on the network connectivity, together with both staleness limits $s\in\{1,5\}$, and three different churn scenarios. Message loss is present throughout all three simulation phases, setup, stabilization, and churn.

\begin{figure}[ht]
  \centering
  \subfloat[Staleness limit $s$=$1$ \label{conn:figure:520,521,522-crop}]{\includegraphics[width=0.98\columnwidth]{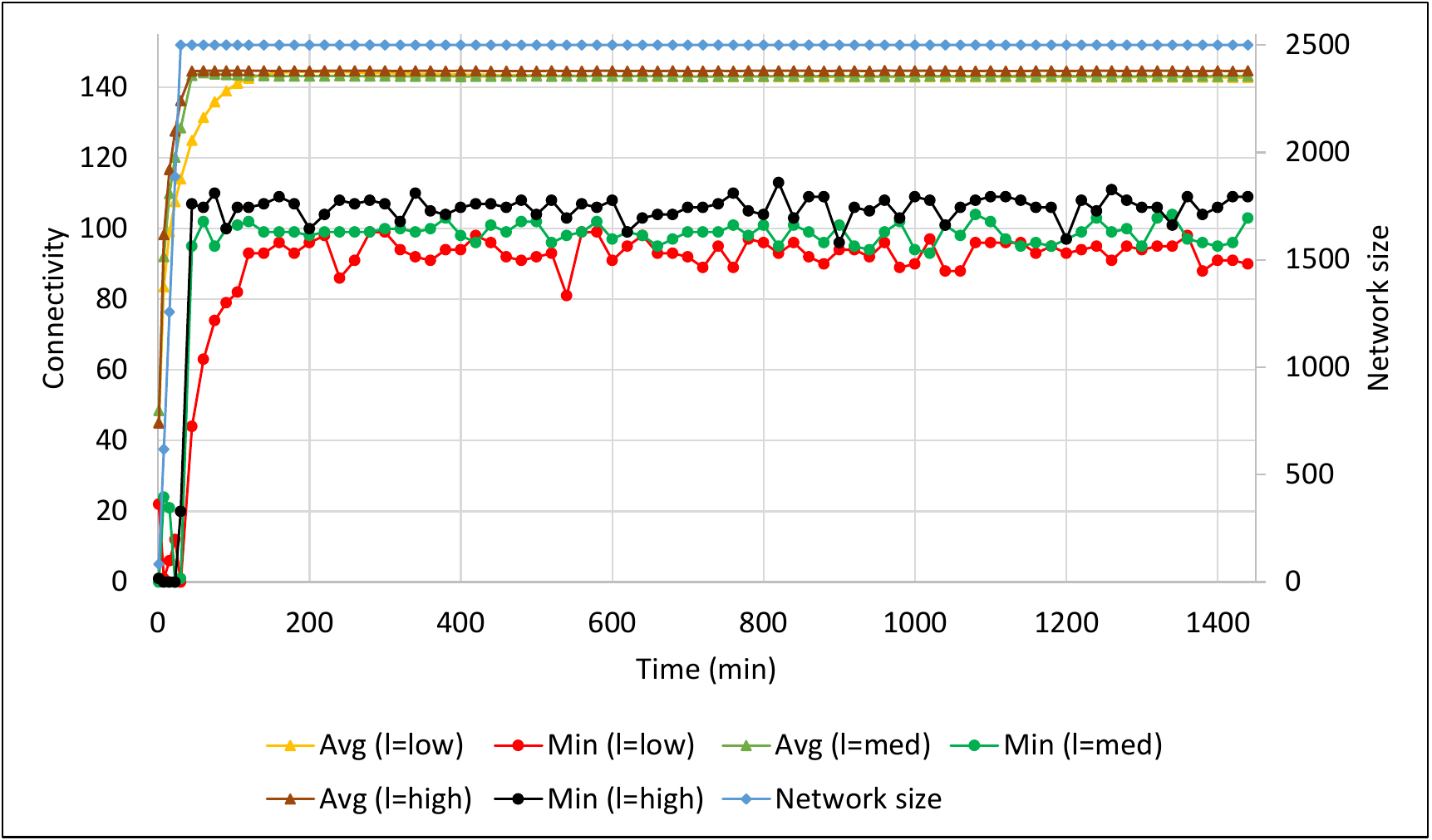}}\qquad
  \subfloat[Staleness limit $s$=$5$ \label{conn:figure:523,524,525-crop}]{\includegraphics[width=0.98\columnwidth]{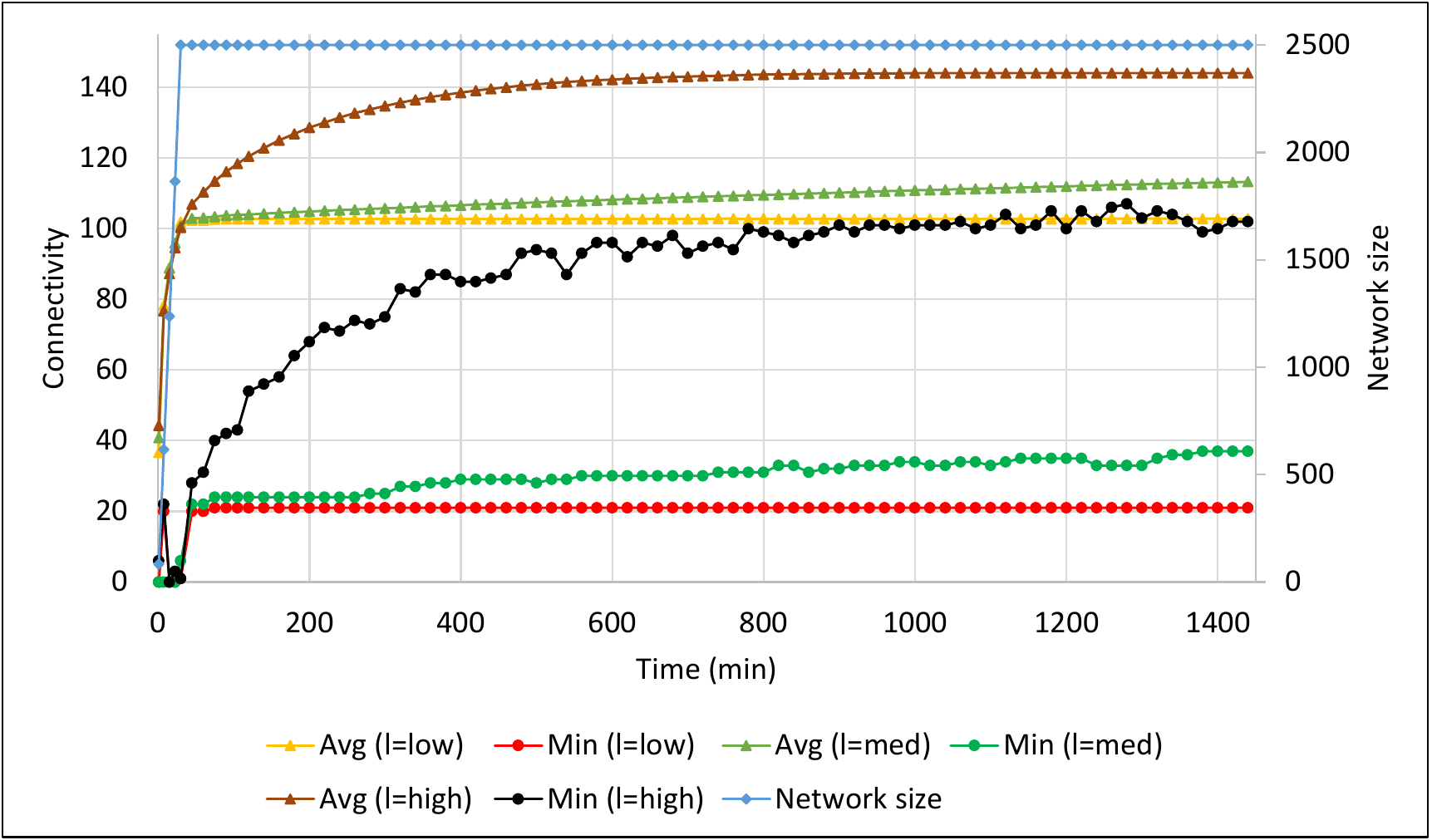}}\qquad
 % \includegraphics[width=\columnwidth]{images/avg_mins_250-crop}\\
  \caption{Simulation\,J, with message loss $l$, $s\in\{1,5\}$, no churn}\label{conn:figure:loss_no_churn}
\end{figure}

Simulation J has no churn.
We present the measurements for this simulation in Figure \ref{conn:figure:loss_no_churn}.
The connectivity during the setup phase is very poor for all loss scenarios and staleness limit values.
Nodes are not able to achieve connectivity immediately on joining the network.
For $s$=$1$ the network shows a quick increase in minimum connectivity immediately after the setup phase.
The minimum connectivity reaches values between 80 and 110, far greater than the bucket size $k$=$20$.
For our three loss scenarios, higher message loss results in higher connectivity.
A similar behavior is visible in simulations A to D with churn 0/1, where nodes leave the network but no new nodes join.
In both cases communication attempts can fail, in simulation J due to message loss, in simulations A to D due to stale nodes.
This leads to the removal of contacts from the routing tables, making room for other contacts.
These results again show, that the structure of a Kademlia network is not ideal regarding connectivity after the network setup or node joins in general.
For $s$=$5$ any structure change due to message loss is much less pronounced than with $s$=$1$, since now a contact is removed from a routing table only after five failing communication attempts in a row, not just one.
The greater staleness limit has a damping effect on both the absolute connectivity and its variance.
Any increase in minimum and average connectivity happens far slower, and the resulting connectivity is lower.
This is especially the case with loss scenarios $medium$ and $low$.
Here, both minimum and average connectivity show a severe decrease compared to $s$=$1$.
For $low$ loss, the positive effect of message loss on minimum connectivity is hardly visible, as the connectivity remains just slightly above $k$=$20$.

We want to remark that, despite its positive effect on connectivity, message loss has of course negative impact on other network aspects, e.g.~the latency or result quality of lookup procedures.
Here, as described in Section \ref{conn:subsec:kademlia}, the termination criterion is either a number of $k$ successfully contacted nodes or a lack of progress.
Message loss can increase the lookup latency, since more communication attempts can be necessary to reach $k$ successfully contacted nodes.
Also, progress may stop earlier because furthering information never reaches the node performing the lookup.

\begin{figure}[ht]
  \centering
  \subfloat[Staleness limit $s$=$1$ \label{conn:figure:526,527,528-crop}]{\includegraphics[width=0.98\columnwidth]{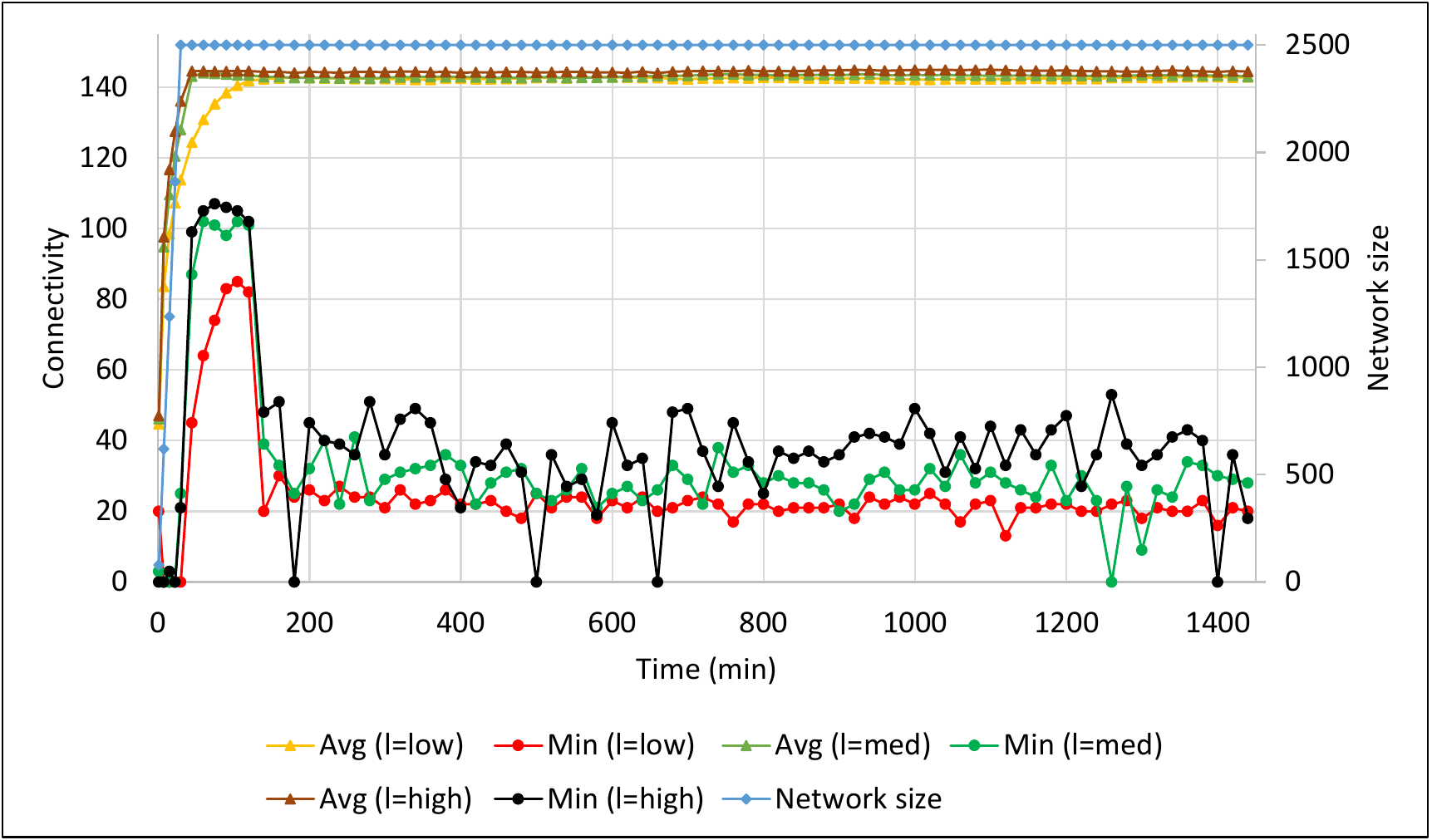}}\qquad
  \subfloat[Staleness limit $s$=$5$ \label{conn:figure:530,531,532-crop}]{\includegraphics[width=0.98\columnwidth]{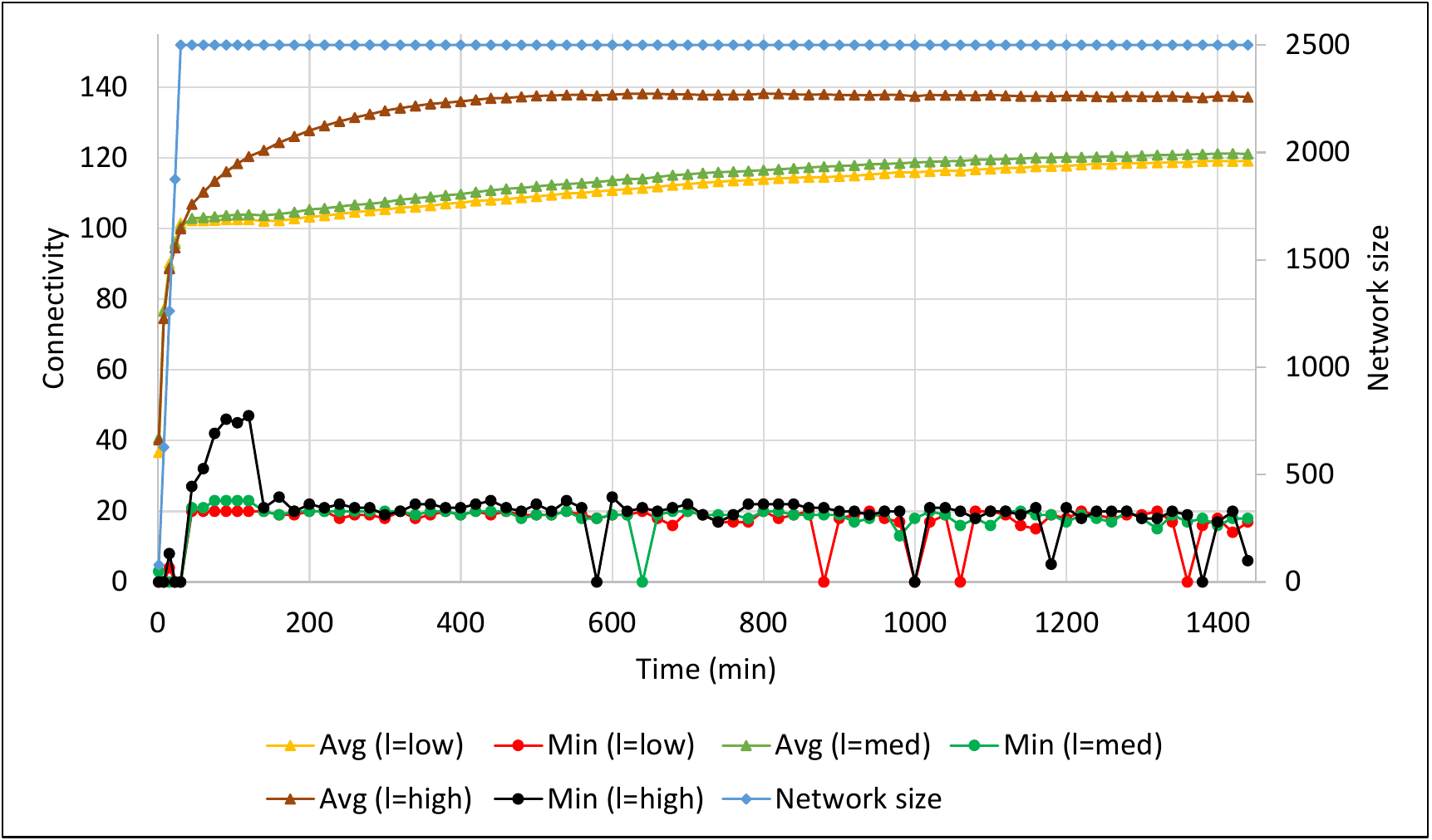}}\qquad
 % \includegraphics[width=\columnwidth]{images/avg_mins_250-crop}\\
  \caption{Simulation\,K, with message loss $l$, $s\in\{1,5\}$, churn 1/1}\label{conn:figure:loss_churn_11}
\end{figure}

In simulation K the churn scenario is 1/1. We show the results in Figure \ref{conn:figure:loss_churn_11}.
For $s$=$1$ the different loss scenarios on average still result in different levels of minimum connectivity during the churn phase.
However, the churn visibly reduces the positive effect of message loss, as those connectivity levels are significantly lower than without churn.
Like in simulation J, a damping effect on connectivity is visible with $s$=$5$. Combined with the churn, it limits the minimum connectivity to about $k$ for all loss scenarios.
Also, the minimum connectivity drops far below $k$ and even down to zero multiple times in the simulation with both staleness limits.
This is due to a small number of nodes not being able to establish connectivity right away in the bootstrap process.

\begin{figure}[ht]
  \centering
  \subfloat[Staleness limit $s$=$1$ \label{conn:figure:533,534,535-crop}]{\includegraphics[width=0.98\columnwidth]{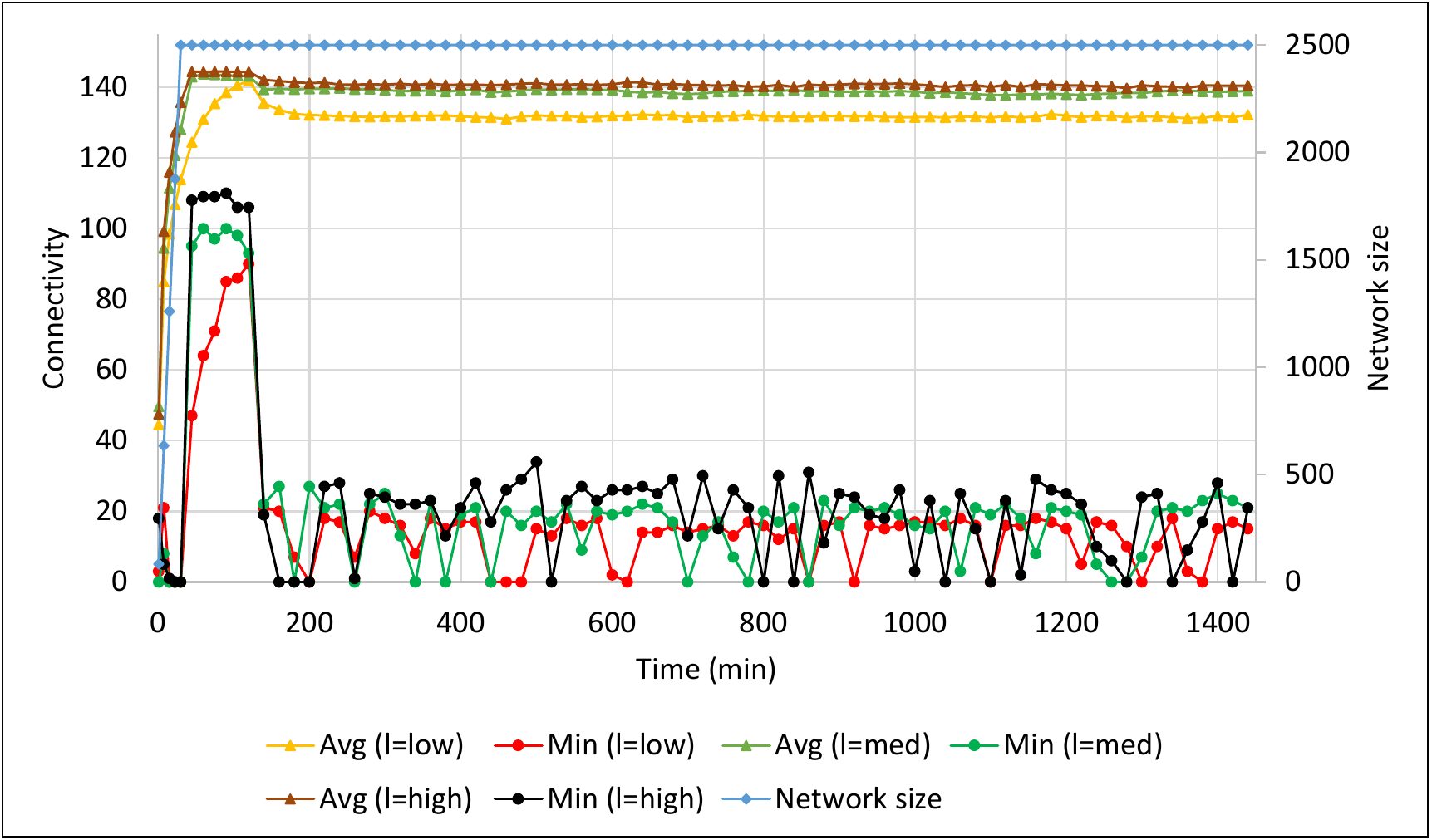}}\qquad
  \subfloat[Staleness limit $s$=$5$ \label{conn:figure:537,538,539-crop}]{\includegraphics[width=0.98\columnwidth]{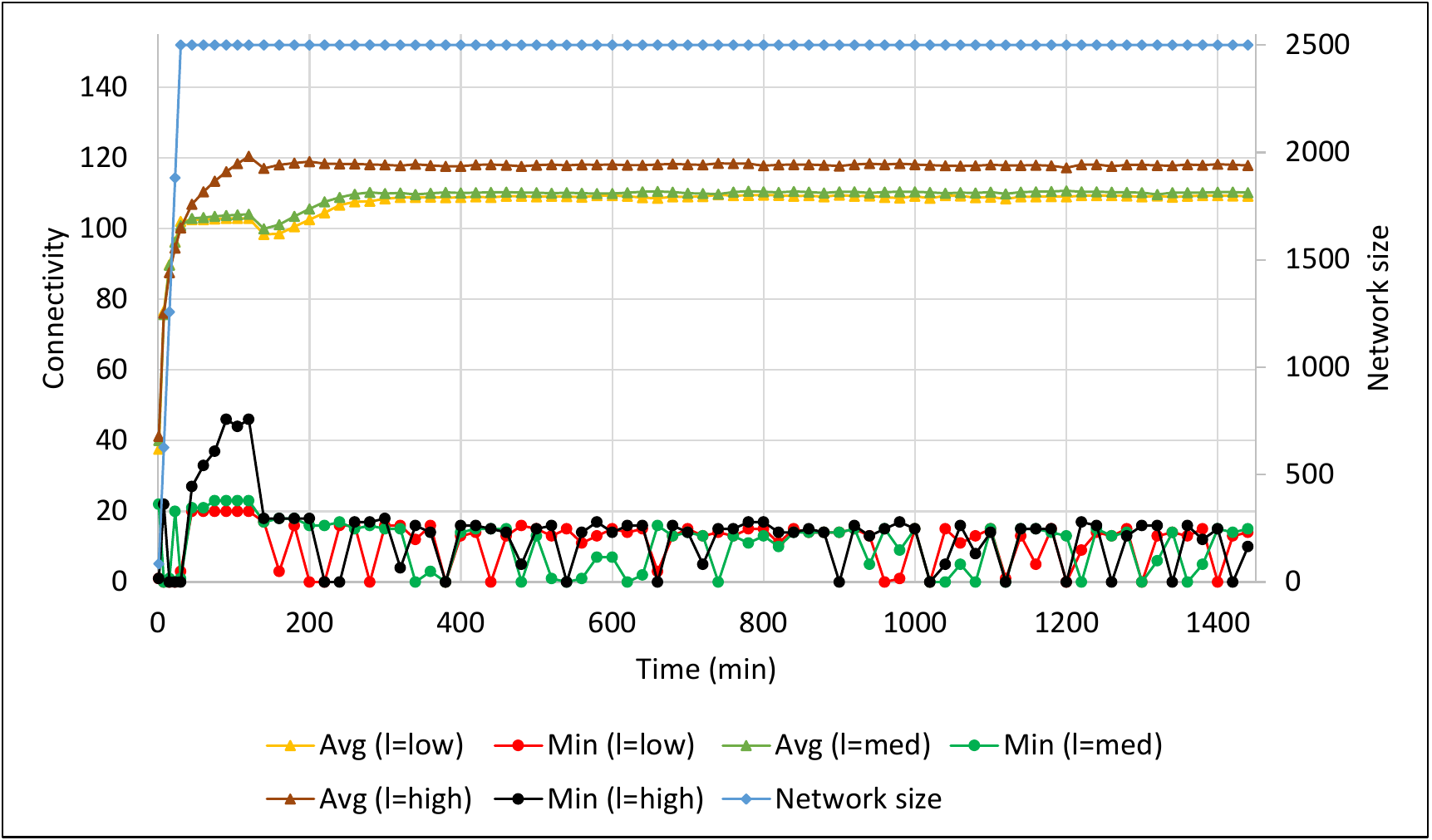}}\qquad
 % \includegraphics[width=\columnwidth]{images/avg_mins_250-crop}\\
  \caption{Simulation\,L, with message loss $l$, $s\in\{1,5\}$, churn 10/10}\label{conn:figure:loss_churn_1010}
\end{figure}

In simulation L the churn scenario is 10/10. We show the results in Figure \ref{conn:figure:loss_churn_1010}.
The stronger churn counters the positive effects of message loss even more, so that now also the average connectivity is affected.
Furthermore, the drops in connectivity due to bootstrap problems are much more frequent.
With the added damping effect from $s$=$5$, the minimum connectivity stays below $k$ at all times during the churn phase.

\section{Conclusion \& Future Work}
\label{conn:sec:conclusion}

In this paper, we analyzed the connectivity of the overlay network Kademlia in multiple simulated scenarios.
We conclude several results from our work.
The network connectivity $\kappa$ of Kademlia strongly correlates with the bucket size $k$.
To achieve a certain resilience level $r$ for an overlay network, we require a network connectivity $\kappa > r$. With our results, we determined that the bucket size needs to be set to a value greater than $r$, i.e., $k>r$. Nevertheless, especially for scenarios with strong churn, the resilience level cannot be guaranteed.
In situations with no or few nodes joining the network, the network connectivity was equal or greater than $k$.
The presence of network traffic greatly enhances the network connectivity, both in terms of absolute connectivity and the time to reach this connectivity.
The effect of 1/1 and 10/10 churn on the network connectivity is ambivalent.
While it can even have a positive effect on the average connectivity, the minimum connectivity drops significantly below $k$ with stronger churn and shows greater variance relative to its mean.
The staleness limit $s$ also has ambivalent effects. While a greater value reduces connectivity variance, it also reduces the overall connectivity level.
Message loss, though generally an undesired property in networks, actually increases the Kademlia network connectivity.

Future work will include investigation of the effects of message loss on the network connectivity. The goal is the development of mechanisms that provide similar connectivity improvements, while avoiding the negative effects of loss. We further plan to extend Kademlia to improve upon the minimum connectivity in all cases and to introduce a parameter to control its connectivity independently of the bucket size.

\section*{Acknowledgment}
The authors thank the German Research Foundation (DFG) for support within the project CYPHOC (WA 2828/1-1).

\bibliographystyle{abbrv}

\bibliography{connection}

\end{document}